\documentclass[preprint,12pt,authoryear]{elsarticle} 

\usepackage{color}
\usepackage{fullpage}
\usepackage{amsmath}
\usepackage{mathrsfs}
\usepackage{subfig}
\usepackage{multirow}
\usepackage{lineno}
\usepackage{caption}
\usepackage{arydshln}
\usepackage{tikz}
\usepackage{pstool}
\usepackage[hidelinks]{hyperref}
\hypersetup{colorlinks,bookmarksopen,linkcolor=blue,citecolor=blue,urlcolor=blue}

\newcommand{\OR}[1]{{#1}}
\newcommand{\bs}[1]{{\boldsymbol{#1}}}
\newcommand{\ignore}[1]{}

\usepackage{amssymb}

\makeatletter
\def\ps@pprintTitle{
 \let\@oddhead\@empty
 \let\@evenhead\@empty
 \def\@oddfoot{\footnotesize\itshape \textcopyright~\the\year. This manuscript version is made available under the \href{https://creativecommons.org/licenses/by-nc-nd/4.0/}{CC-BY-NC-ND 4.0} license.\hfill}
 \let\@evenfoot\@oddfoot}
\makeatother

\begin{document}

\begin{frontmatter}



\title{A Variational Formulation of Dissipative Quasicontinuum Methods\tnoteref{t1}}


\author[affiliation1]{O. Roko\v{s}\corref{mycorrespondingauthor}}
\cortext[mycorrespondingauthor]{Corresponding author.}
\ead{rokosondrej@gmail.com}

\author[affiliation2]{L.A.A. Beex}

\author[affiliation1]{J. Zeman}

\author[affiliation3]{R.H.J. Peerlings}

\address[affiliation1]{Department of Mechanics, Faculty of Civil Engineering, Czech Technical University in Prague, Th\'{a}kurova~7, 166~29 Prague~6, Czech Republic.}

\address[affiliation2]{Facult\'{e} des Sciences, de la Technologie et de la Communication, Campus Kirchberg, Universit\'{e} du Luxembourg~6, rue Richard Coudenhove-Kalergi, L-1359 Luxembourg.}

\address[affiliation3]{Department of Mechanical Engineering, Eindhoven University of Technology, P.O. Box~513, 5600~MB Eindhoven, The Netherlands.}

\tnotetext[t1]{The post-print version of this article published in \emph{International Journal of Solids and Structures}, DOI: 10.1016/j.ijsolstr.2016.10.003.}

\begin{abstract}
Lattice systems and discrete networks with dissipative interactions are successfully employed as meso-scale models of heterogeneous solids. As the application scale generally is much larger than that of the discrete links, physically relevant simulations are computationally expensive. The QuasiContinuum~(QC) method is a multiscale approach that reduces the computational cost of direct numerical simulations by fully resolving complex phenomena only in regions of interest while coarsening elsewhere. In previous work (\citeauthor{BeexDisLatt}, \emph{J. Mech. Phys. Solids}~64, 154--169, 2014), the originally conservative QC methodology was generalized to a virtual-power-based QC approach that includes local dissipative mechanisms. In this contribution, the virtual-power-based QC method is reformulated from a variational point of view, by employing the energy-based variational framework for rate-independent processes (\citeauthor{MieRou:2015}, \emph{Rate-Independent Systems: Theory and Application}, Springer-Verlag, 2015). By construction it is shown that the QC method with dissipative interactions can be expressed as a minimization problem of a properly built energy potential, providing solutions equivalent to those of the virtual-power-based QC formulation. The theoretical considerations are demonstrated on \OR{three} simple examples. For them we verify energy consistency, quantify relative errors in energies, and discuss errors in internal variables obtained for different meshes and two summation rules.
\end{abstract}


\end{frontmatter}


%
%
\section{Introduction}
\label{Sect:1}
Conventional continuum theories discretized by Finite Element~(FE) approaches become problematic at small length-scales and complex material behaviours. In these cases, the underlying microstructure or even the atomistic crystal structure comes into play. This introduces nonlocality, and requires discrete simulations such as structural lattice computations or Molecular Statics~(MS) in order to capture the physics properly. Discrete conservative systems are in their full description conveniently formulated within a variational framework, in which their behaviour follows a minimization of a potential energy~$\mathcal{E}$, i.e.
\begin{equation}
\bs{r}=\underset{\widehat{\bs{r}}\in\mathscr{R}}{\mbox{arg min }}\mathcal{E}(\widehat{\bs{r}}),
\label{Sect:1:Eq:1}
\end{equation}
where~$\widehat{\bs{r}}\in\mathscr{R}$ denotes an arbitrary admissible vector collecting the positions of all lattice atoms (or particles), $\mathscr{R}$ denotes a configuration space, and~$\bs{r} \in \mathscr{R}$ a suitable minimizer, see e.g.~\cite{TadmorModel}, Section~6. For application-scale problems, the construction of~$\mathcal{E}$ and the solution of~\eqref{Sect:1:Eq:1} entails excessive computational efforts because of two facts:
\begin{enumerate}
	\item[\hypertarget{F1}{F1.}] A large number of atoms and bonds contained in fully-resolved systems leads to considerable expenses associated with the solution of the Euler--Lagrange equations involving large-scale energy gradients and Hessians.
	\item[\hypertarget{F2}{F2.}] For the assembly of energies, gradients, and Hessians\footnote{\OR{In MS, it is standard to employ quasi-Newton or completely Hessian-free minimization schemes, requiring only energies and gradients, cf. e.g.~\cite{TadmorModel}, Section~6.2. Contrary to MS, lattice systems are usually solved using a Newton-Raphson scheme that requires also Hessians.}}, all atoms or bonds have to be individually taken into account.
\end{enumerate}
The QuasiContinuum~(QC) methodology, originally formulated by~\cite{Tadmor:1996:QAD}, and extended in various aspects later on, e.g.~\cite{Curtin:2003:ACC,Miller:2002:QCO,Miller:2009:UFP}, overcomes~\hyperlink{F1}{F1} and~\hyperlink{F2}{F2} in two steps. First, \emph{interpolation}, based on a number of selected representative atoms, or \emph{repatoms} for short, constrains the displacements of the remaining atoms in the lattice,
\begin{equation}
\bs{r}=\bs{I}(\bs{r}_\mathrm{rep}),
\label{Sect:1:Eq:2}
\end{equation}
where~$\bs{r}_\mathrm{rep}\in \mathscr{R}_\mathrm{rep}$ stores the positions of all the repatoms, and~$\mathscr{R}_\mathrm{rep}$ denotes a subspace of the original configuration space~$\mathscr{R}$. Because the dimension of~$\mathscr{R}_\mathrm{rep}$ is usually much smaller than that of~$\mathscr{R}$, deficiency~\hyperlink{F1}{F1} is mitigated. The second involves \emph{summation}, in which the energy and governing equations of the reduced model are determined by collecting the contributions only from so-called \emph{sampling atoms}, in analogy to numerical integration of FE method. As a result, an approximation~$\widehat{\mathcal{E}}$ to~$\mathcal{E}$ in~\eqref{Sect:1:Eq:1} is minimized, which resolves~\hyperlink{F2}{F2}. Section~\ref{Sect:3} of this paper presents a more detailed discussion of the two QC approximation steps. Other techniques and further details can be found e.g. in~\cite{TadmorModel,Iyer:QC:2011,Luskin:QC:2013}.

Also at length scales larger than the nanoscale (atomistic length scale), many materials possess discrete underlying structures---regular, irregular, or random---at the micro- or meso-scale; typical representatives are fibrous materials such as paper~\citep{KulachenkoPaper,LiuPaper} or textile~\citep{Potluri:2007:textile}. In such materials the bonds between the fibres (or yarns) take the role of atoms in atomistics. However, since the involved length-scales are larger, the interactions of these "atoms" (i.e. particles) often comprise dissipative processes. Hence, the original QC formulation developed for purely conservative interactions cannot be employed. Initial theoretical developments to lift this limitation have been provided by~\cite{BeexDisLatt,BeexFiber} for fibre plasticity and bond-sliding failure. For the derivation the authors have used a non-variational thermodynamically-consistent framework that employs the following virtual-power statement
\begin{equation}
\dot{\widehat{\bs{r}}}^\mathsf{T}\bs{f}_\mathrm{int}=\dot{\widehat{\bs{r}}}^\mathsf{T}\bs{f}_\mathrm{ext},\quad\forall\dot{\widehat{\bs{r}}}.
\label{Sect:1:Eq:3}
\end{equation}
In Eq.~\eqref{Sect:1:Eq:3}, the dot denotes the derivative with respect to time; the vectors~$\bs{f}_\mathrm{int}$ and~$\bs{f}_\mathrm{ext}$ store components of resulting internal and external forces. This means that the left- and right-hand sides can be identified as the internal and external powers; for further details see~\cite{BeexDisLatt}, Section~2.1. Let us note that in the ideal, smooth and consistent case, the formulation of Eq.~\eqref{Sect:1:Eq:3} would be connected to the one of~\eqref{Sect:1:Eq:1} via the relation~$\bs{f}_\mathrm{int} - \bs{f}_\mathrm{ext} = \partial\mathcal{E}(\widehat{\bs{r}})/\partial\widehat{\bs{r}}$. Throughout this paper, the QC approach based on Eq.~\eqref{Sect:1:Eq:3} will be referred to as the \emph{virtual-power-based QC}. The virtual-power-based QC framework has been employed in various contexts and proven to be efficient while accurate, see e.g.~\cite{BeexDisLatt}. However, from this formulation, it is not entirely clear whether the governing equations derived from Eq.~\eqref{Sect:1:Eq:3} are also energetically consistent; it may happen that some terms are missing, cf. e.g.~\cite{VarLoc}, Tab.~1, for an example in continuum gradient plasticity. Variational approaches may furthermore be considered to provide finer information about system evolution such as microstructure pattern formation or phase transition, see e.g.~\cite{Ortiz1999:nonconvex}, \cite{Carstensen2002:nonconvex}, and~\cite{Schrder:2013}. In the case of adaptivity, better error estimates and mesh refinement capabilities for localized phenomena \OR{(such as damage)} can be explored \OR{in highly nonlinear problems}, \OR{cf. e.g.~\cite{Radovitzky:1999}}. \OR{From a broader perspective, the variational formulation offers a consistent framework convenient for, e.g., the rigorous treatment of evolutions that exhibit discontinuities in time, investigations of structural stability using energy landscapes arising from time-incremental minimization, or direct employment of non-linear optimization algorithms.} Finally, the variational formulation allows us to extend the conservative QC methodology to an entire class of rate-independent internal mechanisms in a natural way.\footnote{\OR{In principle, extensions to inertial and viscous effects are possible as well. For the sake of simplicity and clarity, any rate effects are omitted throughout this contribution, and the interested reader is referred to~\cite{MieRou:2015}, Chapter~5 and references therein.}}

The goal of this paper is therefore to reformulate the virtual-power-based QC framework for internal dissipative processes in terms of variational principles and show that the obtained solutions \emph{coincide} for both formulations in the case of plasticity with isotropic hardening. To that end, a suitable potential~$\Pi$ will be constructed such that 
\begin{equation}
\bs{q}\in\underset{\widehat{\bs{q}}\in\mathscr{Q}}{\mbox{arg min }}\Pi(\widehat{\bs{q}}),
\label{Sect:1:Eq:4}
\end{equation}
describing the state of the system in analogy to~\eqref{Sect:1:Eq:1}. Here, however, $\bs{q}$ denotes a general state variable that also includes internal dissipative variables. Furthermore, $\mathscr{Q}$ is an abstract state space, and the inclusion sign~$\in$ indicates that the potential~$\Pi$ is generally nonsmooth or may have multiple minima. In analogy to standard QC, a reduced variable~$\bs{q}_\mathrm{red}\in\mathscr{Q}_\mathrm{red}$ and an approximate energy~$\widehat{\Pi}$ will be introduced in order to alleviate~\hyperlink{F1}{F1} and~\hyperlink{F2}{F2}. In what follows, the approach based on Eq.~\eqref{Sect:1:Eq:4} will be referred to as the \emph{variational QC}. Its construction falsifies the statement presented in~\cite{BeexDisLatt}, Section~1, claiming that the solution to Eq.~\eqref{Sect:1:Eq:3} \emph{cannot} be obtained by direct minimization of an energy potential. 

In order to construct the full energy potential~$\Pi$, we employ the variational formulation of rate-independent processes as introduced in an \emph{abstract setting}  by~\cite{MieRou:2015} that is closely related to applications in continuum mechanics. Earlier studies were provided e.g. by~\cite{Francfort1993stable}, \cite{HanReddy}, \cite{Francfort:1998}, \cite{Ortiz:CMAME:1999}, \cite{Charlotte}, \cite{Hackl:PRSA:2008}, \cite{Conti:JMPS:2008}, and~\cite{Kochmann:CMT:2010}. Section~\ref{Sect:2} of this paper first briefly introduces definitions and basic principles of the theory. Second, the approach is reformulated in the particular context of discrete lattice systems.

The governing equations associated with~\eqref{Sect:1:Eq:4} will be addressed in Section~\ref{Sect:4}, where we recall the Alternating Minimization~(AM) method, see also~\cite{BourdinVar}. Since the energy potential~$\Pi$ for plasticity is nonsmooth, we will also briefly discuss the return-mapping algorithm suitable for its minimization.

Before closing this contribution by a summary and conclusions in Section~\ref{Sect:6}, we perform numerical tests on \OR{three} benchmark examples presented in Section~\ref{Sect:5}, \OR{two of which} have been adopted from~\cite{BeexDisLatt}, Section~4, and~\cite{BeexHO}, Section~4.2. We demonstrate that both approaches, represented by Eqs.~\eqref{Sect:1:Eq:3} and~\eqref{Sect:1:Eq:4}, lead to energetically-consistent solutions for the exact and central summation rules presented in~\cite{BeexQC} and~\cite{BeexCSumRule}. \OR{The third example then  presents both global as well as local quantities for an indentation test. Finally,} we show that despite the significant dimension reduction and time savings achieved by the QC method, the obtained errors in stored and dissipated energies due to interpolation and summation are rather low: the relative errors in energies do not exceed~$4\,\%$, while the simulation time is decreased by a factor of 4~-- 30 depending on the triangulation, loading, and geometry.
%
%
\section{Rate-Independent Variational Plasticity}
\label{Sect:2}
%
%
%
\subsection{General Considerations}
\label{SubSect:2.1}
The variational formulation for rate-independent processes comprises several steps and relies on two principles~\eqref{S} and~\eqref{E}, which are described below (for details see~\citealp{Mielke2002,RateMie,Mielke:CMAME:2004,MieRou:2015}). The state of the system within a fixed time horizon~$[0,T]$ is described in terms of a non-dissipative variable~$\bs{r}(t)\in\mathscr{R}$, and a dissipative component~$\bs{z}(t)\in\mathscr{Z}$. The latter specifies all irreversible processes at time~$t\in[0,T]$, where~$t$ denotes a pseudo-time parametrizing the quasi-static evolution process. The state of the system is fully characterized by the state variable~$\bs{q}(t)=(\bs{r}(t),\bs{z}(t))\in\mathscr{Q}=\mathscr{R}\times\mathscr{Z}$. Furthermore, we consider the total free (Helmholtz type) energy of the system~$\mathcal{E} : [0,T]\times\mathscr{Q}\rightarrow\mathbb{R}$ together with the dissipation distance~$\mathcal{D}(\bs{z}_2,\bs{z}_1)$, $\mathcal{D}:\mathscr{Z}\times\mathscr{Z}\rightarrow\mathbb{R}^+\cup\{+\infty\}$ which specifies the minimum amount of energy spent by the continuous transition between two consecutive states~$\bs{z}_1$ and~$\bs{z}_2$; a rigorous definition and further details on~$\mathcal{D}$ can be found in~\cite{MieRou:2015}, Section~3.2. Then, the process~$\bs{q}:[0,T]\rightarrow\mathscr{Q}$ is called an energetic solution to the initial-value problem described by~$(\mathcal{E},\mathcal{D},\bs{q}_0)$ if it satisfies the following two principles~\eqref{S} and~\eqref{E}, together with an initial condition~\eqref{I}:
\begin{enumerate}
	
	\item[(S)] {\bf Global stability:} for all~$t\in[0,T]$ and for all $\widehat{\bs{q}}\in\mathscr{Q}$
	\begin{equation}
	\mathcal{E}(t,\bs{q}(t))\leq\mathcal{E}(t,\widehat{\bs{q}})+\mathcal{D}(\widehat{\bs{z}},\bs{z}(t)),
	\tag{S}
	\label{S}
	\end{equation}
	which requires the solution to be the global minimum of the sum~$\mathcal{E}+\mathcal{D}$. To see this recall that the definition of the global minimum of~$\mathcal{E}+\mathcal{D}$ reads
	\begin{equation}
		\mathcal{E}(t,\bs{q}(t)) + \mathcal{D}(\bs{z}(t),\bs{z}(s)) \leq \mathcal{E}(t,\widehat{\bs{q}}) + \mathcal{D}(\widehat{\bs{z}},\bs{z}(s)),
		\label{min}
	\end{equation}
	where~$\bs{q}(s) \in \mathscr{Q}$ denotes previous configuration at time~$s \leq t$. As~$\mathcal{D}$ is an extended-quasidistance function (for definition see e.g.~\citealt{MieRou:2015}, Section~2.1.1), it satisfies the triangle inequality~$\mathcal{D}(\widehat{\bs{z}},\bs{z}(s)) \leq \mathcal{D}(\bs{z}(t),\bs{z}(s)) + \mathcal{D}(\widehat{\bs{z}},\bs{z}(t))$, which used in~\eqref{min} provides~\eqref{S}.

	\item[(E)] {\bf Energy equality:} for all $t\in[0,T]$
	\begin{equation}
	\mathcal{E}(t,\bs{q}(t))+\mathrm{Var}_{\mathcal{D}}(\bs{q};0,t)=\mathcal{E}(0,\bs{q}(0))+\int_0^t\mathcal{P}(s)\,\mbox{d}s,
	\tag{E}
	\label{E}
	\end{equation}
	which expresses the energy balance in terms of the internal energy, the dissipated energy~$\mathrm{Var}_{\mathcal{D}}$, and the time-integrated power of external forces~$\mathcal{P}$.
	
	\item[(I)] {\bf Initial condition:}
	\begin{equation}
	\bs{q}(0)=\bs{q}_0.
	\tag{I}
	\label{I}
	\end{equation}
\end{enumerate}
In the second principle~\eqref{E}, the dissipation along a curve~$\bs{q}$ is expressed as
\begin{equation}
\mathrm{Var}_{\mathcal{D}}(\bs{q};0,t) =  \sup\left\{\sum_{k=1}^{n}\mathcal{D}(\bs{z}(t_{k}),\bs{z}(t_{k-1}))\right\},
\label{SubSect:2.1:Eq:1}
\end{equation}
where the supremum is taken over all~$n\in\mathbb{N}$ and all partitions of the time interval~$[0,t]$, $0=t_0<t_1<\dots<t_{n}=t$. Introducing a time discretization of~$[0,T]$, the two principles~\eqref{S} and~\eqref{E} along with initial condition~\eqref{I} naturally give rise to an
\begin{itemize}
	\item[(IP)] {\bf Incremental problem:} for $k=1,\ldots,n_T$
	\begin{equation}
	\bs{q}(t_k)\in\underset{\widehat{\bs{q}}\in\mathscr{Q}}{\mbox{arg min }}
	\Pi^k(\widehat{\bs{q}};\bs{q}(t_{k-1}))
	\tag{IP}
	\label{IP}
	\end{equation}
\end{itemize}
amenable to a numerical solution in which each step is realized as a minimization problem of an incremental energy
\begin{equation}
\Pi^k(\widehat{\bs{q}};\bs{q}(t_{k-1}))=\mathcal{E}(t_k,\widehat{\bs{q}})+\mathcal{D}(\widehat{\bs{z}},\bs{z}(t_{k-1})).
\label{IE}
\end{equation}
The main conceptual difficulty with~\eqref{IP} is that it represents a global minimization problem, which is computationally cumbersome for non-convex energies. It is reasonable, however, to assume that stable solutions of~\eqref{IP} are associated with local minima. Then, multiple minimizers can exist, but only the ones satisfying~\eqref{E} are adopted as proper solutions. It has been shown that as long as the solution~$\bs{q}(t)$ remains continuous in time, the energy balance holds (see~\citealt{StabPhaMarMaur,Pham:2013}). In the examples in Section~\ref{Sect:5} we verify that this is indeed the case for hardening plasticity, thereby justifying the local minimization approach adopted in this work. As a result, the energy balance~\eqref{E} need not be enforced explicitly. 
%
%
\subsection{Lattices with Dissipative Internal Processes}
\label{SubSect:DissLatt}
\begin{figure}
	\centering
	\subfloat[configurations and kinematics]{\includegraphics[scale=1]{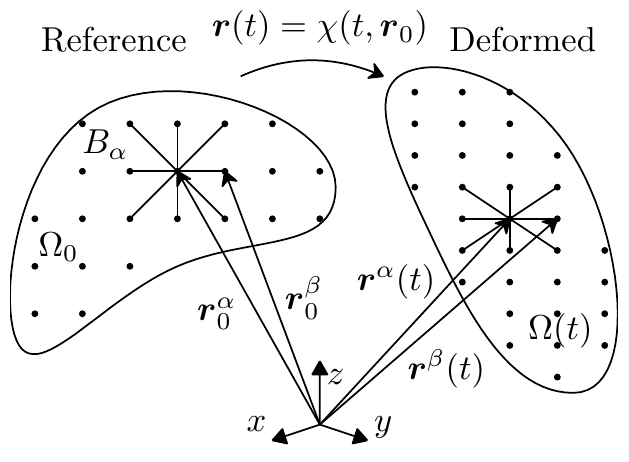}\label{DissLatt:Fig:1a}}
	\hspace{2em}
	\subfloat[single bond]{\includegraphics[scale=1]{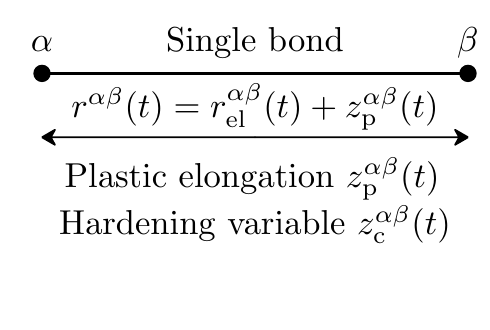}\label{DissLatt:Fig:1b}}
	\caption{Sketch of~(a) geometric variables and two system configurations, reference~$\Omega_0$ and current~$\Omega(t)$, and~(b) single bond setup.}
	\label{DissLatt:Fig:1}
\end{figure}

To specify the above formulation for lattice systems, we start by introducing a geometrical setting and necessary notation (depicted in Fig.~\ref{DissLatt:Fig:1}), followed by explicit definitions of energies. Before doing so, let us mention that the term "atoms" is meant to represent individual particles or nodes of the underlying (non-atomistic) lattice, consistently with the original QC methodology developed for atomistic systems. Furthermore, for clarity we confine our exposition to two spatial dimensions; the extension to three dimensions is straightforward.
%
%
\subsubsection{Geometry and Internal Variables}
\label{SubSubSect:2.2.1}
The domain~$\Omega_0 \subset \mathbb{R}^{2}$ in a reference configuration contains a set~$N_\mathrm{ato}$ of~$n_\mathrm{ato}=\#N_\mathrm{ato}$ atoms, where~$\#\bullet$ returns the cardinality of a set~$\bullet$. The reference spatial position of each atom~$\alpha\in N_\mathrm{ato}$ is specified by a vector~$\bs{r}_0^\alpha\in\mathbb{R}^{2}$, and can be expressed as a linear combination of primitive vectors in analogy to the Bravais lattices since we confine our attention to regular structures only.\footnote{For the sake of clarity, we consider only nearest-neighbour interactions in what follows, although the presented theory can be easily generalized to long-range or multi-body interactions. For extensions to beam structures see e.g.~\cite{BeexBeams}.} All positions~$\bs{r}_0^\alpha$ are collected in a column matrix~$\bs{r}_0=[\bs{r}_0^1,\dots,\bs{r}_0^{n_\mathrm{ato}}]^\mathsf{T}$, $\bs{r}_0 \in \mathbb{R}^{2\,n_\mathrm{ato}}$, for convenience. Note that throughout this paper, Greek indices refer to atom numbers whereas Latin indices are reserved for spatial coordinates or other integer parametrizations. Each atom~$\alpha$ is further attributed with a set~$B_\alpha \subset N_\mathrm{ato}$ of its nearest neighbours; recall that for the truss structures of interest, the Verlet list---i.e. the lists of all the neighbours~$B_\alpha$ for all atoms~$\alpha \in N_\mathrm{ato}$---does not change in time. The distance between two atoms~$\alpha$ and~$\beta$ and the list of all inter-atomic distances in the reference configuration are denoted as
\begin{subequations}
	\label{SubSect:DissLatt:Eq:1}
	\begin{align}
	r_0^{\alpha\beta}(\bs{r}_0)&=||\bs{r}_0^\beta-\bs{r}_0^\alpha||_2,\label{SubSect:DissLatt:Eq:1a}\\
	\{r_0^{\alpha\beta}(\bs{r}_0)\}&=\{r_0^{\alpha\beta}\,|\,\alpha=1,\dots,n_\mathrm{ato},\ \beta \in B_\alpha,\mbox{ duplicity removed}\},\label{SubSect:DissLatt:Eq:1b}
	\end{align}
\end{subequations}
where~$||\bullet||_2$ represents the Euclidean norm. Since~$r_0^{\alpha\beta} = r_0^{\beta\alpha}$, the set~$\{r_0^{\alpha\beta}\}$ in~\eqref{SubSect:DissLatt:Eq:1b} consists of~$n_\mathrm{bon}$ components, where~$n_\mathrm{bon}$ is the number of all the bonds of the system collected in a set~$N_\mathrm{bon}$, i.e.~$\# N_\mathrm{bon} = n_\mathrm{bon}$. Throughout this paper we employ the symbol~$\alpha\beta$ in two contexts: in the context of atoms, $r_0^{\alpha\beta}$ means the distance in the reference configuration between two atoms~$\alpha,\beta \in N_\mathrm{ato}$ (as used above), whereas in the context of bonds the same symbol~$r_0^{\alpha\beta}$ means the length of the $p$-th bond in the reference configuration, $p = \alpha\beta$, $p \in N_\mathrm{bon}$, with end atoms~$\alpha,\beta \in N_\mathrm{ato}$; a similar convention applies to other physical quantities.

Deformation mapping~$\chi(t,\bs{r}_0)$ transforms~$\Omega_0$ from the reference to the current configuration~$\Omega(t)\subset\mathbb{R}^{2}$, where the locations of the atoms are specified by position vectors~$\bs{r}^\alpha(t)$, $\alpha=1,\dots,n_\mathrm{ato}$. In analogy to~$\bs{r}_0$, we collect all~$\bs{r}^\alpha$ in a column matrix~$\bs{r}(t)=[\bs{r}^1(t),\dots,\bs{r}^{n_\mathrm{ato}}(t)]^\mathsf{T}$, $\bs{r}(t) \in \mathbb{R}^{2\,n_\mathrm{ato}}$, that represents also the abstract non-dissipative variable\footnote{Strictly speaking, the non-dissipative component should consist of~$\bs{r}_\mathrm{el}^{\alpha\beta}$ for appropriate~$\alpha$ and~$\beta$, where~$\bs{r}_\mathrm{el}^{\alpha\beta}=\bs{r}^{\alpha\beta}-\frac{\bs{r}^{\alpha\beta}}{r^{\alpha\beta}}z_\mathrm{p}^{\alpha\beta}$. Because such an affine transformation does not affect the results, we adopt~$\bs{r}^{\alpha\beta}$ instead of~$\bs{r}_\mathrm{el}^{\alpha\beta}$ as our primal variables from now on for convenience.}. Furthermore, we introduce the distance measure between two atoms~$r^{\alpha\beta}(\bs{r}(t))$, and the set of all distances~$\{r^{\alpha\beta}(\bs{r}(t))\}$, cf. Eq.~\eqref{SubSect:DissLatt:Eq:1}. Due to the time-dependent Dirichlet boundary conditions and possible kinematic constraints, $\mathscr{R}(t)$ is a function of time and it forms a manifold in~$\mathbb{R}^{2\,n_\mathrm{ato}}$. 

Each bond is further endowed with two internal variables: the plastic slip or elongation of the bond~$z_\mathrm{p}^{\alpha\beta}(t)$, and the cumulative plastic slip (a hardening variable) $z_\mathrm{c}^{\alpha\beta}(t)$. For clarity, we introduce column matrices~$\bs{z}_\mathrm{p}(t)$ and~$\bs{z}_\mathrm{c}(t)$, both in~$\mathbb{R}^{n_\mathrm{bon}}$, collecting~$z_\mathrm{p}^{\alpha\beta}(t)$ and~$z_\mathrm{c}^{\alpha\beta}(t)$ of all bonds. Then, the abstract dissipative variable can be specified as~$\bs{z}(t)=(\bs{z}_\mathrm{p}(t),\bs{z}_\mathrm{c}(t))$, and~$\mathscr{Z}$ as~$\mathbb{R}^{2\,n_\mathrm{bon}}$ (recall that~$\bs{z}(t) \in \mathscr{Z}$).
%
%
\subsubsection{Definition of Energies}
\label{SubSubSect:2.2.2}
Having described the physical variables, we proceed to the formal definitions of the energies. Because no external forces will be used, the elastic part of the incremental energy is not an explicit function of time~$t$ and reads as
\begin{equation}
\mathcal{E}(\widehat{\bs{q}})=\mathcal{V}_\mathrm{int}(\widehat{\bs{r}},\widehat{\bs{z}}_\mathrm{p})+\mathcal{V}_\mathrm{hard}(\widehat{\bs{z}}_\mathrm{c}).
\label{SubSect:DissLatt:Eq:2}
\end{equation}
Note that hatted variables~$\widehat{\bullet}$ represent an arbitrary admissible configuration of the system, whereas the absence of hats indicates that these state variables are minimizers of~\eqref{IP}, see also Eq.~\eqref{IE}. The interatomic potential energy~$\mathcal{V}_\mathrm{int}$ in~\eqref{SubSect:DissLatt:Eq:2} specifies a recoverable part of the energy stored in all atom interactions. It is sufficient to adhere to pair potentials, for which
\begin{equation}
\mathcal{V}_\mathrm{int}(\widehat{\bs{r}},\widehat{\bs{z}}_\mathrm{p})=\mathcal{V}_\mathrm{int}(\{\widehat{r}^{\alpha\beta}(\bs{r})\},\widehat{\bs{z}}_\mathrm{p})=\frac{1}{2}\sum_{\alpha,\beta\in B_\alpha}\phi^{\alpha\beta}(\widehat{r}^{\alpha\beta},\widehat{z}^{\alpha\beta}_\mathrm{p}),
\label{SubSect:DissLatt:Eq:3}
\end{equation}
where the first equality holds as a consequence of the principle of potential invariance. Note that permutation symmetry requires~$\phi^{\alpha\beta}=\phi^{\beta\alpha}$, and that~$\phi^{\alpha\beta}$ corresponds to the elastic portion of the energy stored in a single bond stretched to a length~$\widehat{r}^{\alpha\beta}$ with a plastic elongation~$\widehat{z}_\mathrm{p}^{\alpha\beta}$. Further,
\begin{equation}
\mathcal{V}_\mathrm{hard}(\widehat{\bs{z}}_\mathrm{c})=\frac{1}{2}\sum_{\alpha,\beta\in B_\alpha}h^{\alpha\beta}(\widehat{z}^{\alpha\beta}_\mathrm{c})
\label{SubSect:DissLatt:Eq:4}
\end{equation}
reflects an unrecoverable part of the stored energy, locked in all bonds due to hardening effects, where~$h^{\alpha\beta}(\bullet)$ denotes the hardening pair potential of a single bond. The dissipation distance for a single bond, $\mathcal{D}^{\alpha\beta}$, between two different states~$\widehat{\bs{z}}_1$ and~$\widehat{\bs{z}}_2$ is defined as
\begin{equation}
\mathcal{D}^{\alpha\beta}(\widehat{\bs{z}}_2,\widehat{\bs{z}}_1)=
\left\{
\begin{aligned}
& f_0^{\alpha\beta}|\widehat{z}_{\mathrm{p},2}^{\alpha\beta}-\widehat{z}_{\mathrm{p},1}^{\alpha\beta}|&&\mbox{ if }\widehat{z}_{\mathrm{c},2}^{\alpha\beta}\geq \widehat{z}_{\mathrm{c},1}^{\alpha\beta}+|\widehat{z}_{\mathrm{p},2}^{\alpha\beta}-\widehat{z}_{\mathrm{p},1}^{\alpha\beta}|\\
&+\infty && \mbox{ otherwise,}
\end{aligned}
\right.\quad\alpha\beta=1,\dots,n_\mathrm{bon},
\label{SubSect:DissLatt:Eq:5}
\end{equation}
where~$f_0^{\alpha\beta}>0$ is an initial yield force. The total dissipation distance then collects the contributions of all bonds, namely
\begin{equation}
\mathcal{D}(\widehat{\bs{z}}_2,\widehat{\bs{z}}_1)=\frac{1}{2}\sum_{\alpha,\beta\in B_\alpha}\mathcal{D}^{\alpha\beta}(\widehat{\bs{z}}_2,\widehat{\bs{z}}_1).
\label{SubSect:DissLatt:Eq:6}
\end{equation}
Note that setting~$\mathcal{V}_\mathrm{hard}=\mathcal{D}=0$ and~$\bs{z}(t) = \bs{0}$ reduces~\eqref{IP} to standard MS defined in Eq.~\eqref{Sect:1:Eq:1}; for further details we refer the interested reader to, e.g., \cite{TadmorModel}, Chapter~6.

Let us close this section with the observation that the total incremental energy~$\Pi^k$ can be expressed in two equivalent forms: as a sum over all atom sites such as described above, or as a sum over all bonds. This allows us, therefore, to introduce the incremental bond, $\widetilde{\pi}^k_{\alpha\beta}$, and site, $\widehat{\pi}^k_\alpha$, energies of the form
\begin{subequations}
	\label{SubSect:DissLatt:Eq:7}
	\begin{align}
	\widetilde{\pi}_{\alpha\beta}^k(\widehat{\bs{q}};\bs{q}(t_{k-1})) & = \phi^{\alpha\beta}(\widehat{r}^{\alpha\beta},\widehat{z}_\mathrm{p}^{\alpha\beta})+h^{\alpha\beta}(\widehat{z}^{\alpha\beta}_\mathrm{c})+\mathcal{D}^{\alpha\beta}(\widehat{\bs{z}},\bs{z}(t_{k-1})),\ {\alpha\beta}=1,\dots,n_\mathrm{bon},\label{SubSect:DissLatt:Eq:7a}\\
	\widehat{\pi}_\alpha^k(\widehat{\bs{q}};\bs{q}(t_{k-1})) & =\frac{1}{2}\sum_{\beta\in B_\alpha}\widetilde{\pi}_{\alpha\beta}^k(\widehat{\bs{q}},\bs{q}(t_{k-1})),\ \alpha=1,\dots,n_\mathrm{ato},\label{SubSect:DissLatt:Eq:7b}
	\end{align}
\end{subequations}
and write
\begin{equation}
\Pi^k(\widehat{\bs{q}};\bs{q}(t_{k-1})) = \sum_{\alpha\beta=1}^{n_\mathrm{bon}}\widetilde{\pi}_{\alpha\beta}^k(\widehat{\bs{q}};\bs{q}(t_{k-1})) = \sum_{\alpha=1}^{n_\mathrm{ato}}\widehat{\pi}_\alpha^k(\widehat{\bs{q}};\bs{q}(t_{k-1})).
\label{SubSect:DissLatt:Eq:8}
\end{equation}
The reason for introducing two equivalent expressions for~$\Pi^k$ is that the site energies~\eqref{SubSect:DissLatt:Eq:7b} are convenient for the minimization of~\eqref{IP} with respect to the kinematic variable~$\widehat{\bs{r}}$, whereas the bond energies~\eqref{SubSect:DissLatt:Eq:7a} are more suitable for the minimization with respect to the internal variable~$\widehat{\bs{z}}$.
%
%
\section{Quasicontinuum Methodology}
\label{Sect:3}
Let us proceed to the two QC steps introduced to mitigate excessive computational demands implied by~\hyperlink{F1}{F1} and~\hyperlink{F2}{F2} that are inherently associated with the minimization problems~\eqref{Sect:1:Eq:1} and~\eqref{IP} for conservative and dissipative systems. In the two sections below, on interpolation and summation, we explain how these steps apply on the incremental energy~$\Pi^k$.
%
%
\subsection{Interpolation}
\label{SubSect:3.1}
Upon specifying a subset of atoms~$N_\mathrm{rep}^\mathrm{ato}\subseteq N_\mathrm{ato}$, $\# N_\mathrm{rep}^\mathrm{ato}=n_\mathrm{rep}\ll n_\mathrm{ato}$ that determine the deformation state of the system, one can reconstruct the positions of all the remaining atoms through interpolation:
\begin{equation}
\widehat{\bs{r}}=\bs{\Phi}\widehat{\bs{r}}_\mathrm{rep},
\label{SubSect:3.1:Eq:1}
\end{equation}
cf. Eq.~\eqref{Sect:1:Eq:2}. Here, $\widehat{\bs{r}}_\mathrm{rep}\in\mathscr{R}_\mathrm{rep}(t)$ represents a column matrix of all representative atoms' position vectors for any admissible configuration, and the matrix~$\boldsymbol{\Phi}$ stores the basis vectors spanning~$\mathscr{R}_\mathrm{rep}(t)$ by columns. Analogously, $\bs{r}_{0,\mathrm{rep}}$ represents a vector of repatoms' positions in the reference configuration. Eq.~\eqref{SubSect:3.1:Eq:1} basically introduces a geometric equality constraint for~$\widehat{\bs{r}}$ which, upon substitution into Eq.~\eqref{IE}, entails that the incremental energy becomes a function of~$\widehat{\bs{r}}_\mathrm{rep}$ and~$\bs{r}_\mathrm{rep}(t_{k-1})$, i.e.
\begin{equation}
\Pi^k(\widehat{\bs{r}},\widehat{\bs{z}}_\mathrm{p},\widehat{\bs{z}}_\mathrm{c};\bs{r}(t_{k-1}),\bs{z}_\mathrm{p}(t_{k-1}),\bs{z}_\mathrm{c}(t_{k-1}))=\Pi^k(\bs{\Phi}\widehat{\bs{r}}_\mathrm{rep},\widehat{\bs{z}}_\mathrm{p},\widehat{\bs{z}}_\mathrm{c};\bs{\Phi}\bs{r}_\mathrm{rep}(t_{k-1}),\bs{z}_\mathrm{p}(t_{k-1}),\bs{z}_\mathrm{c}(t_{k-1})),
\label{SubSect:3.1:Eq:3}
\end{equation}
reducing the number of degrees of freedom associated with the kinematic state of the system from~$2\,n_\mathrm{ato}$ to~$2\,n_\mathrm{rep}$, which is substantial if~$n_\mathrm{rep} \ll n_\mathrm{ato}$. Minimization in~\eqref{IP} with respect to~$\widehat{\bs{r}}\in\mathscr{R}(t)$ then changes to a minimization over the subspace~$\mathscr{R}_\mathrm{rep}(t)$ which, due to the linearity in~\eqref{SubSect:3.1:Eq:1}, effectively yields a projection of the full solution to that subspace. In order to specify~$\bs{\Phi}$, one usually introduces a triangulation of~$\Omega_0$ equipped with piecewise-affine shape functions with compact support (in analogy to the FE methodology, where the nodes are the repatoms). This triangulation is finely resolved (to fully recover the underlying lattice) in the region of interest and coarsely elsewhere. Evaluations of these shape functions at the positions of all atoms then provide the base vectors,
\begin{equation}
\bs{\Phi}=\Phi_{(2\alpha-1)(2j-1)}=\Phi_{(2\alpha)(2j)}=\varphi_{\beta_j}(\bs{r}_0^\alpha),\ \alpha\in N_\mathrm{ato},\ \beta_j\in N_\mathrm{rep}^\mathrm{ato},\ j=1,\dots,n_\mathrm{rep}^\mathrm{ato},
\label{SubSect:3.1:Eq:5}
\end{equation}
where~$\beta_j=N_\mathrm{rep}^\mathrm{ato}(j)$ denotes the $j$-th element of the set~$N_\mathrm{rep}^\mathrm{ato}$, and~$\varphi_{\beta}(\bs{r}_0^\alpha)$ represents a shape function associated with a repatom~$\beta$ that is evaluated for an atom~$\alpha$ in the undeformed configuration. In addition, we assume that the shape functions satisfy the Kronecker-delta property, i.e.~$\varphi_\beta(\bs{r}_0^\gamma)=\delta^{\beta\gamma}$ for~$\beta,\gamma\in N_\mathrm{rep}^\mathrm{ato}$, as well as partition-of-unity property, i.e.~$\sum_{\beta\in N_\mathrm{rep}^{\mathrm{ato}}}\varphi_\beta(\bs{r}_0^\alpha)=1$ for each~$\alpha\in N_\mathrm{ato}$. Note that higher-order or meshless interpolations are possible as well, see e.g.~\cite{Xiao:IJNME:2007}, \cite{Kwon:JCP:2009}, \cite{BeexHO}, or~\cite{YangMMM}.

Similarly, we could introduce internal variables associated with repbonds~$N_\mathrm{rep}^\mathrm{bon}\subseteq N_\mathrm{bon}$, and through interpolation represent all the remaining internal variables of the system,
\begin{equation}
\widehat{\bs{z}}_\bullet=\bs{\Psi}\widehat{\bs{z}}_{\bullet,\mathrm{rep}},
\label{SubSect:3.1:Eq:2}
\end{equation}
where~$\bullet$ stands either for~"p" or~"c". Eq.~\eqref{SubSect:3.1:Eq:2} then would constrain the admissible vector of either plastic elongations~$\widehat{\bs{z}}_\mathrm{p}$ or cumulative plastic elongations~$\widehat{\bs{z}}_\mathrm{c}$. This approach would, however, lead to a different technique than introduced by~\cite{BeexDisLatt}, with possible benefits in error estimation, cf. e.g.~\cite{ChenROM}, since the procedure treats all unknowns equally and resembles a structure-preserving Reduced-Order-Modelling method, cf. e.g.~\cite{ROM} or~\cite{Fritzen:CMAME:2014}. Some similarities with the models for bond-sliding by~\cite{BeexFiber} can be observed. However, let us emphasize that in their case, the internal variables associated with sliding are attributed to nodes rather than to bonds, so that they can be approximated in the same way as the displacements. Within the virtual-power-based QC framework for lattices with plasticity, only a subset of unknowns (denoted~$\widehat{\bs{z}}_{\bullet,\mathrm{sam}}$) is sought as well, but this is an assumption of the \emph{summation} rule, yielding an approximation to incremental energy~$\Pi^k$ rather than a geometrical constraint, see Section~\ref{SubSect:3.2} below. Equation~\eqref{SubSect:3.1:Eq:2} is therefore not included in further considerations and is left as a possible future challenge.
%
%
\subsection{Summation}
\label{SubSect:3.2}
\begin{figure}
	\centering
	\includegraphics[scale=1]{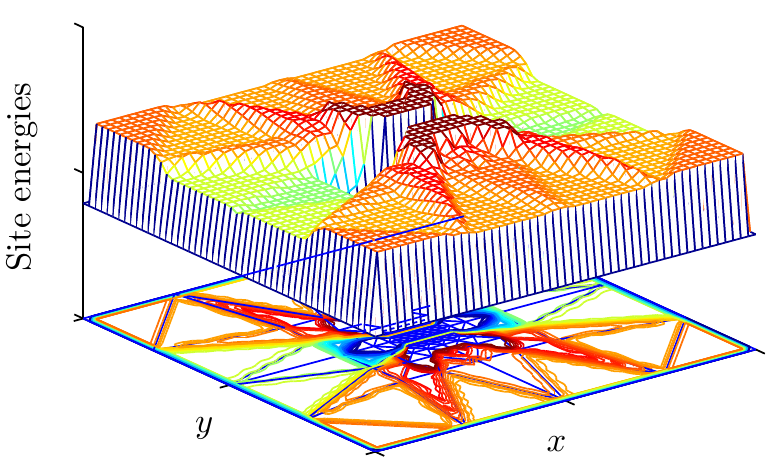}
	\caption{Site energies of a simple QC system uniformly stretched along $x$-axis with an inhomogeneity in the central region; energy peaks near the inhomogeneity are trimmed for visualisation purposes. In the $x$-$y$ plane below the surface, contour lines are indicated for better clarity.}
	\label{SubSect:3.2:Fig:1}
\end{figure}

The interpolation step with piece-wise affine shape functions ensures that all atomic bonds inside triangles follow a homogeneous deformation, and therefore yields constant site energies for atoms that have all their nearest neighbours inside the same triangle, cf. Fig.~\ref{SubSect:3.2:Fig:1}. This observation was significantly used during the construction of summation rules briefly recalled below.

Let us express the approximation to the incremental energy defined in Eqs.~\eqref{IE} and~\eqref{SubSect:DissLatt:Eq:8} as
\begin{equation}
\Pi^k\approx\widehat{\Pi}^k = \bs{\Sigma}^\mathsf{T}\widehat{\bs{\pi}},
\label{SubSect:3.2:Eq:1}
\end{equation}
where~$\bs{\Sigma}$ in general has the form~$\bs{\Sigma}=[w_1,\dots,w_{n_\mathrm{ato}}]^\mathsf{T}$, $w_\alpha\geq 0$, and~$\widehat{\bs{\pi}}=[\widehat{\pi}_1^k,\dots,\widehat{\pi}_{n_\mathrm{ato}}^k]^\mathsf{T}$ stores all individual incremental site energies. Choosing weight factors~$w_\alpha=1$, $\alpha=1,\dots,n_\mathrm{ato}$, clearly recovers the full sum presented in~\eqref{SubSect:DissLatt:Eq:8}. Introducing a set of sampling atoms~$S_\mathrm{ato}$ such that~$w_\alpha > 0$ for~$\alpha\in S_\mathrm{ato}$ and~$w_\beta=0$ for~$\beta\in N_\mathrm{ato}\backslash S_\mathrm{ato}$, $\# S_\mathrm{ato}=n_\mathrm{sam}^\mathrm{ato}\ll n_\mathrm{ato}$, one can rewrite~\eqref{SubSect:3.2:Eq:1} also as
\begin{equation}
\widehat{\Pi}^k = \sum_{\alpha\in S_\mathrm{ato}}w_\alpha\widehat{\pi}_\alpha^k.
\label{SubSect:3.2:Eq:2}
\end{equation}
Analogously to sampling atoms, we introduce a set of sampling bonds, $S_\mathrm{bon}$, $\# S_\mathrm{bon}=n_\mathrm{sam}^\mathrm{bon}$, defined as those bonds that are connected to all the sampling atoms~$S_\mathrm{ato}$ with removed duplicity, recall Eq.~\eqref{SubSect:DissLatt:Eq:1b}. The approximation of the incremental energy can then be expressed again as a sum over all sampling atoms (Eq.~\eqref{SubSect:3.2:Eq:2}) or as a sum over all sampling bonds (in analogy to Eq.~\eqref{SubSect:DissLatt:Eq:8}). Consequently, the dimensionality of the internal variable reduces from~$2\, n_\mathrm{bon}$ to~$2\, n_\mathrm{sam}^\mathrm{bon}$, because all variables associated with~$N_\mathrm{bon} \backslash S_\mathrm{bon}$ become irrelevant. The reduced dissipative variable is denoted as~$\bs{z}_\mathrm{sam}(t)=(\bs{z}_{\mathrm{p},\mathrm{sam}}(t),\bs{z}_{\mathrm{c},\mathrm{sam}}(t))\in\mathscr{Z}_\mathrm{sam}$, where~$\mathscr{Z}_\mathrm{sam}$ is identified with~$\mathbb{R}^{2\, n_\mathrm{sam}^\mathrm{bon}}$.

Let us recall Eq.~\eqref{SubSect:3.1:Eq:2} and the discussion related to it. From that, it may be clear that a mapping from~$\bs{z}_\mathrm{sam}$ back to the full solution~$\bs{z}$ (i.e.~$\bs{z}_\mathrm{sam} \rightarrow \bs{z}$ in analogy to Eq.~\eqref{SubSect:3.1:Eq:2}) is not necessarily unique. The reduction of the global state variable from~$\bs{q}(t)=(\bs{r}(t),\bs{z}(t))\in\mathscr{Q}$ to~$\bs{q}_\mathrm{red}(t)=(\bs{r}_\mathrm{rep}(t),\bs{z}_\mathrm{sam}(t))\in\mathscr{Q}_\mathrm{red}=\mathscr{R}_\mathrm{rep}(t)\times\mathscr{Z}_\mathrm{sam}$ is thus formed by the combination of the two QC steps that cannot be separated.

\begin{figure}
	\centering
	\subfloat[exact]{\includegraphics[scale=1]{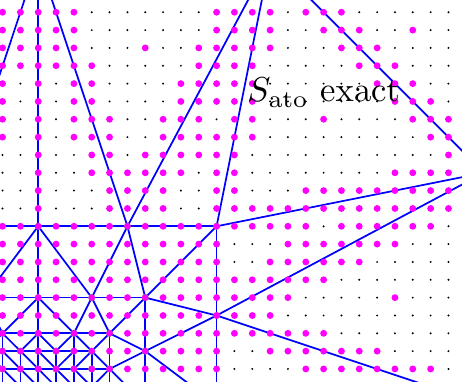}\label{SubSect:3.2:Fig:2a}}
	\hspace{2em}
	\subfloat[central]{\includegraphics[scale=1]{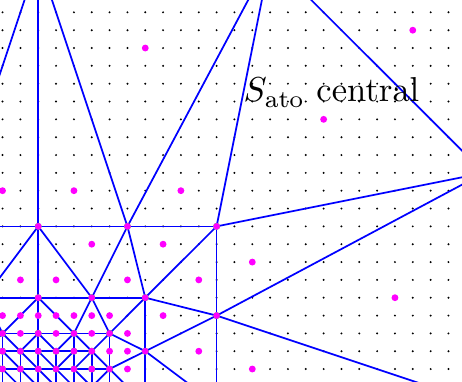}\label{SubSect:3.2:Fig:2b}}
	\caption{Schematic representation of the two sets of sampling atoms contained in~$S_\mathrm{ato}$; small black dots denote ordinary atom sites and larger magenta dots emphasize the sampling atoms. The triangulation is depicted in the blue colour.}
	\label{SubSect:3.2:Fig:2}
\end{figure}

In order to compute the total energy exactly (i.e. to integrate the site energy function in Fig.~\ref{SubSect:3.2:Fig:1} exactly), it is necessary to incorporate all variations along the triangles' edges. For a large triangle, the energy of only one atom site multiplied by the number of atoms within that triangle suffices to represent the plateau, whereas the atoms near the edges and in the fully resolved region must all be taken into account explicitly. This yields the so-called \emph{exact summation rule} proposed by~\cite{BeexQC}, in which the explicit procedure how to compute weight factors $w_\alpha$ ($\alpha \in S_\mathrm{ato}$) and how to determine the set of sampling atoms~$S_\mathrm{ato}$ can be found. Another sampling scheme focuses on the centres of the triangles and is therefore referred to as \emph{central summation rule}, as only the plateaus and vertex atoms are sampled. See~\cite{BeexCSumRule} or~\cite{Amelang:2015} for further details and Fig.~\ref{SubSect:3.2:Fig:2} for a schematic representation of the two summation rules. Compared to the exact summation rule, the central summation rule is cheaper, but introduces approximation errors.
%
%
\section{Numerical Solution Strategies}
\label{Sect:4}
Directly approaching~\eqref{IP} with respect to both variables~$\widehat{\bs{r}}$ and~$\widehat{\bs{z}}$ can be cumbersome. Instead, it is relatively straightforward to perform the minimization in a staggered way, i.e. with respect to kinematic and internal variables sequentially, resulting in the AM method. This scheme appears in the literature also under names such as Block-Nonlinear Gauss--Seidel Method or Block-Coordinate Descent Method. For the original formulation the reader is referred to~\cite{CsizarAM}, for recent works to e.g.~\cite{AM:Byrne}, and for applications in variational fracture to e.g.~\cite{BourdinVar}.

In the context of the variational QC method, the AM procedure needs to be applied to the incremental problem~\eqref{IP} with the incremental energy defined by Eq.~\eqref{IE}. For a fixed time step~$t_k$, it results in the following scheme
\ignore{The AM procedure applied to the variational QC framework presented in this contribution, i.e. solving problem~\eqref{IP} with incremental energy defined in Eq.~\eqref{IE} For a fixed time step~$t_k$, results in the following scheme}
\begin{enumerate}[(i)]
	\item {\bf Initialization:} $\bs{r}^0 = \bs{r}(t_{k-1})$, $\bs{z}^0_\mathrm{p} = \bs{z}_\mathrm{p}(t_{k-1})$, and~$\bs{z}^0_\mathrm{c} = \bs{z}_\mathrm{c}(t_{k-1})$
	\item {\bf General iteration:} $l=0,1,\dots,\mbox{until convergence}$
	\begin{subequations}
		\label{AM}
		\begin{align}
		\bs{r}^{l+1} & = \underset{\widehat{\bs{r}}\in\mathscr{R}(t_k)}{\mbox{arg min }}\Pi^k(\widehat{\bs{r}},\bs{z}_\mathrm{p}^{l},\bs{z}_\mathrm{c}^{l};\bs{q}(t_{k-1})),\label{AMa}\tag{AMa}\\
		\bs{z}_\mathrm{p}^{l+1} & \in \underset{\widehat{\bs{z}}_\mathrm{p}\in\mathbb{R}^{n_\mathrm{bon}}}{\mbox{arg min }}\Pi^k(\bs{r}^{l+1},\widehat{\bs{z}}_\mathrm{p},\bs{z}_\mathrm{c}^{l};\bs{q}(t_{k-1})),\label{AMb}\tag{AMb}\\
		\bs{z}_\mathrm{c}^{l+1} & \in \underset{\widehat{\bs{z}}_\mathrm{c}\in\mathbb{R}^{n_\mathrm{bon}}}{\mbox{arg min }}\Pi^k(\bs{r}^{l+1},\bs{z}_\mathrm{p}^{l+1},\widehat{\bs{z}}_\mathrm{c};\bs{q}(t_{k-1})).\label{AMc}\tag{AMc}
		\end{align}
	\end{subequations}
\end{enumerate}
In the first time increment, i.e. for~$k = 1$, the initial condition~\eqref{I} is used. Let us note that a similar scheme applies also to more general definitions of the incremental energy~$\Pi^k$ applicable to multi-body potentials or long-range interactions.
%
%
\subsection{Full-Lattice Computation}
\label{SubSect:4.1}
For hardening plasticity, i.e. for~$\frac{\mathrm{d}h^{\alpha\beta}(\widehat{z}_\mathrm{c})}{\mathrm{d}\widehat{z}_\mathrm{c}}>0$ and~$\frac{\mathrm{d}^2h^{\alpha\beta}(\widehat{z}_\mathrm{c})}{\mathrm{d}\widehat{z}_\mathrm{c}^2}>0$, the minimization in~\eqref{AMc} can be performed in closed form. For arbitrary time step~$t_k$ and iteration~$l$, the solution reads
\ignore{Assuming~$\frac{\mathrm{d}h^{\alpha\beta}(\widehat{z}_\mathrm{c})}{\mathrm{d}\widehat{z}_\mathrm{c}}>0$ and~$\frac{\mathrm{d}^2h^{\alpha\beta}(\widehat{z}_\mathrm{c})}{\mathrm{d}\widehat{z}_\mathrm{c}^2}>0$ (i.e. assuming hardening plasticity) first~$\Pi^k$ can be minimized with respect to~$\widehat{\bs{z}}_\mathrm{c}$ according to~\eqref{AMc}. This provides the following explicit relation for time step~$t_k$ and iteration~$l$:}
\begin{equation}
z^{\alpha\beta,l+1}_\mathrm{c}=z^{\alpha\beta}_\mathrm{c}(t_{k-1})+|z^{\alpha\beta,l+1}_\mathrm{p}-z^{\alpha\beta}_\mathrm{p}(t_{k-1})|,\ {\alpha\beta}=1,\dots,n_\mathrm{bon}.
\label{SubSect:4.1:Eq:1}
\end{equation}
This relation follows from the bond-wise formulation of the energy in Eqs.~\eqref{SubSect:DissLatt:Eq:7a} and~\eqref{SubSect:DissLatt:Eq:8}, and from the definition of the dissipation distance in Eq.~\eqref{SubSect:DissLatt:Eq:5}. The minimization then decomposes into independent problems
\begin{equation}
z_\mathrm{c}^{\alpha\beta,l+1}\in\min_{\widehat{z}_\mathrm{c}^{\alpha\beta}\in\mathbb{R}}\widetilde{\pi}_{\alpha\beta}^k(r^{\alpha\beta,l+1},z^{\alpha\beta,l+1}_\mathrm{p},\widehat{z}_\mathrm{c}^{\alpha\beta};\bs{q}(t_{k-1})),\ {\alpha\beta}=1,\dots,n_\mathrm{bon}
\label{SubSect:4.1:Eq:1.1}
\end{equation}
solved by~\eqref{SubSect:4.1:Eq:1}. Substituting~\eqref{SubSect:4.1:Eq:1} into definition~\eqref{IE} provides the \emph{reduced} incremental energy, cf. e.g.~\cite{MieRou:2015}, Section~3.1.2, or~\cite{Carstensen2002:nonconvex},
\begin{equation}
\Pi_\mathrm{red}^k(\widehat{\bs{r}},\widehat{\bs{z}}_\mathrm{p};\bs{q}(t_{k-1})),
\label{Sect:4:Eq:2}
\end{equation}
and the AM algorithm simplifies to steps~\eqref{AMa}~-- \eqref{AMb} for~$\Pi_\mathrm{red}^k$.

Continuing with step~\eqref{AMa}, $\Pi_\mathrm{red}^k$ is sufficiently smooth with respect to~$\widehat{\bs{r}}$, so that the standard Newton's algorithm can be employed. In what follows, two nested iteration cycles will be used where~$l$ relates to AM and~$i$ to Newton's algorithm. Assuming that~$\widehat{\bs{z}}_\mathrm{p}=\bs{z}_\mathrm{p}^l$ and, $t_k$ and~$l$ are fixed in~\eqref{Sect:4:Eq:2}, the second-order Taylor expansion in~$\widehat{\bs{r}}$, in the vicinity of~$\widehat{\bs{r}}^i$, is applied to obtain the following stationarity conditions
\begin{equation}
\bs{K}^i(\widehat{\bs{r}}^{i+1}-\widehat{\bs{r}}^i)+\bs{f}^i = \bs{0},
\label{Sect:4:Eq:5}
\end{equation}
where
\begin{subequations}
\label{Sect:4:Eq:4}
\begin{align}
\bs{f}^i &= \bs{f}(\widehat{\bs{r}}^i) = \left.\frac{\partial\Pi_\mathrm{red}^k(\widehat{\bs{r}},\bs{z}_\mathrm{p}^l;\bs{q}(t_{k-1}))}{\partial\widehat{\bs{r}}}\right|_{\widehat{\bs{r}} = \widehat{\bs{r}}^i},\\
\bs{K}^i &= \bs{K}(\widehat{\bs{r}}^i)=\left.\frac{\partial^2\Pi_\mathrm{red}^k(\widehat{\bs{r}},\bs{z}_\mathrm{p}^l;\bs{q}(t_{k-1}))}{\partial\widehat{\bs{r}}\partial\widehat{\bs{r}}}\right|_{\widehat{\bs{r}} = \widehat{\bs{r}}^i}.
\end{align}
\end{subequations}
Condition~\eqref{Sect:4:Eq:5} supplies a system of linear equations for increments~$\widehat{\bs{r}}^{i+1}-\widehat{\bs{r}}^i$. Iterating~\eqref{Sect:4:Eq:5} and~\eqref{Sect:4:Eq:4} until convergence of~$||\bs{f}^i||_2$ then yields~$\bs{r}^{l+1}$. As usual, Dirichlet boundary conditions are imposed through known increments for constrained atoms. For tying conditions, the constrained primal-dual minimization procedure is applied, cf. Section~\ref{Sect:5}. We refer also to \cite{TadmorModel}, Section~6.4.4, for similar approaches used for MS systems. The gradients~$\bs{f}^i$ and Hessians~$\bs{K}^i$ are provided in detail in Eqs.~\eqref{Sect:A:Eq:1}~-- \eqref{Sect:A:Eq:4} in~\ref{Sect:A} for the reader's convenience.

Before dealing with the non-smooth step~\eqref{AMb}, the energy is again rewritten into the bond-wise form, cf. Eqs.~\eqref{SubSect:DissLatt:Eq:7a} and~\eqref{SubSect:DissLatt:Eq:8}. Consequently, \eqref{AMb} can be treated in analogy to~\eqref{SubSect:4.1:Eq:1.1}, i.e.
\begin{equation}
\begin{aligned}
z_\mathrm{p}^{\alpha\beta,l+1}&\in\min_{\widehat{z}_\mathrm{p}^{\alpha\beta}\in\mathbb{R}}\widetilde{\pi}^k_{\mathrm{red},\alpha\beta}(r^{\alpha\beta,l+1},\widehat{z}^{\alpha\beta}_\mathrm{p};\bs{q}(t_{k-1}))\Longleftrightarrow\\
0&\in\partial\,\widetilde{\pi}^k_{\mathrm{red},\alpha\beta}(r^{\alpha\beta,l+1},z^{\alpha\beta,l+1}_\mathrm{p};\bs{q}(t_{k-1})),\ \alpha\beta=1,\dots,n_\mathrm{bon},
\end{aligned}
\label{Sect:4:Eq:9}
\end{equation}
where~$\partial$ denotes the subdifferential with respect to~$\widehat{z}^{\alpha\beta}_\mathrm{p}$, cf. e.g.~\citep{RoubicekNonlin} or~\citep[Section~8.1]{BonnansOptim}, and~$\widetilde{\pi}^k_{\mathrm{red},\alpha\beta}$ represents a reduced bond energy. In mechanics terms, Eq.~\eqref{Sect:4:Eq:9} represents the Karush--Kuhn--Tucker complementarity conditions of a stretched uniform bar with isotropic hardening, and can thus be solved with the standard return-mapping algorithm described in detail e.g. in~\cite{SimoInelast}, Section~1.4.2. 
\ignore{Note that in some special cases, problem~\eqref{Sect:4:Eq:9} can even be solved explicitly.}
%
%
\subsection{QC Computation}
\label{SubSect:4.2}
Instead of minimizing the exact incremental energy~$\Pi_\mathrm{red}^k$, its approximation~$\widehat{\Pi}_\mathrm{red}^k$ in terms of the reduced variable~$\bs{q}_\mathrm{red} \in \mathscr{Q}_\mathrm{red}$ is minimized in the variational QC formulation. Using the chain rule in the Taylor series expansion applied in step~\eqref{AMa}, recall Eqs.~\eqref{SubSect:3.1:Eq:3} and~\eqref{SubSect:3.2:Eq:2}, provides the stationarity conditions
\begin{equation}
\bs{H}^i(\widehat{\bs{r}}^{i+1}_\mathrm{rep}-\widehat{\bs{r}}^i_\mathrm{rep}) + \bs{G}^i = \bs{0},
\label{SubSect:4.2:Eq:1}
\end{equation}
with
\begin{subequations}
\label{SubSect:4.2:Eq:2}
\begin{align}
\bs{G}^i& = \bs{G}(\widehat{\bs{r}}^i_\mathrm{rep}) = \left.\bs{\Phi}^\mathsf{T}\frac{\partial\widehat{\Pi}_\mathrm{red}^k(\widehat{\bs{r}},\bs{z}_\mathrm{p}^l;\bs{q}_\mathrm{red}(t_{k-1}))}{\partial\widehat{\bs{r}}}\right|_{\widehat{\bs{r}}=\bs{\Phi}\widehat{\bs{r}}_\mathrm{rep}^i},\\
\bs{H}^i& = \bs{H}(\widehat{\bs{r}}^i_\mathrm{rep}) = \left.\bs{\Phi}^\mathsf{T}\frac{\partial^2\widehat{\Pi}_\mathrm{red}^k(\widehat{\bs{r}},\bs{z}_\mathrm{p}^l;\bs{q}_\mathrm{red}(t_{k-1}))}{\partial\widehat{\bs{r}}\partial\widehat{\bs{r}}}\bs{\Phi}\right|_{\widehat{\bs{r}} = \bs{\Phi}\widehat{\bs{r}}_\mathrm{rep}^i},
\end{align}
\end{subequations}
where the partial derivatives are expressed as
\begin{subequations}
\label{SubSect:4.2:Eq:3}
\begin{align}
\frac{\partial\widehat{\Pi}_\mathrm{red}^k(\widehat{\bs{r}},\bs{z}_\mathrm{p}^l;\bs{q}_\mathrm{red}(t_{k-1}))}{\partial\widehat{\bs{r}}}&= \sum_{\alpha\in S_\mathrm{ato}}w_\alpha\frac{\partial\widehat{\pi}^k_{\mathrm{red},\alpha}(\widehat{\bs{r}},\bs{z}_\mathrm{p}^l;\bs{q}_\mathrm{red}(t_{k-1}))}{\partial\widehat{\bs{r}}}=\sum_{\alpha\in S_\mathrm{ato}}w_\alpha\bs{f}^\alpha_\mathrm{int}(\widehat{\bs{r}}),\\
\frac{\partial^2\widehat{\Pi}_\mathrm{red}^k(\widehat{\bs{r}},\bs{z}_\mathrm{p}^l;\bs{q}_\mathrm{red}(t_{k-1}))}{\partial\widehat{\bs{r}}\partial\widehat{\bs{r}}}&=\sum_{\alpha\in S_\mathrm{ato}}w_\alpha\frac{\partial^2\widehat{\pi}^k_{\mathrm{red},\alpha}(\widehat{\bs{r}},\bs{z}_\mathrm{p}^l;\bs{q}_\mathrm{red}(t_{k-1}))}{\partial\widehat{\bs{r}}\partial\widehat{\bs{r}}}=\sum_{\alpha\in S_\mathrm{ato}}w_\alpha\bs{K}^\alpha(\widehat{\bs{r}}).
\end{align}
\end{subequations}
For definitions and explicit expressions of~$\bs{f}^\alpha_\mathrm{int}$ and~$\bs{K}^\alpha$ see Eqs.~\eqref{Sect:A:Eq:1} and~\eqref{Sect:A:Eq:3}. The converged solution of~\eqref{SubSect:4.2:Eq:1} is denoted by~$\bs{r}^{l+1}_\mathrm{rep}$. 

In order to minimize in~\eqref{AMb}, again the bond-wise version of the approximate incremental energy is employed, yielding
\begin{equation}
\begin{aligned}
z_\mathrm{p}^{\alpha\beta,l+1}&\in\min_{\widehat{z}_\mathrm{p}^{\alpha\beta}\in\mathbb{R}}\overline{w}_{\alpha\beta}\widetilde{\pi}^k_{\mathrm{red},\alpha\beta}(r^{\alpha\beta,l+1},\widehat{z}^{\alpha\beta}_\mathrm{p};\bs{q}_\mathrm{red}(t_{k-1}))\Longleftrightarrow\\
0&\in\partial\,\overline{w}_{\alpha\beta}\widetilde{\pi}^k_{\mathrm{red},\alpha\beta}(r^{\alpha\beta,l+1},z^{\alpha\beta,l+1}_\mathrm{p};\bs{q}_\mathrm{red}(t_{k-1})),\ {\alpha\beta}\in S_\mathrm{bon},
\end{aligned}
\label{SubSect:4.2:Eq:4}
\end{equation}
where~$\overline{w}_{\alpha\beta},\alpha\beta\in S_\mathrm{bon},$ denotes the weight factor associated with a bond. Because the problems in~\eqref{SubSect:4.2:Eq:4} are independent, the weights~$\overline{w}_{\alpha\beta}$ are irrelevant (though they can be easily established from~$w_\alpha$) and the problem can be solved sequentially using the return-mapping algorithm in analogy to Eq.~\eqref{Sect:4:Eq:9}, but only over a subset of bonds~$S_\mathrm{bon}\subseteq N_\mathrm{bon}$. 

Interestingly, the resulting governing equations \emph{exactly} coincide with those provided by~\cite{BeexDisLatt}, indicating that the virtual-power-based formulation is also variationally-consistent. Let us note that the introduction of constraints for the dissipative variables, as outlined in Eq.~\eqref{SubSect:3.1:Eq:2}, would lead to coupled systems of equations for~$\bs{z}_{\mathrm{p},\mathrm{rep}}$ and a different (non-local) minimization strategy would be needed instead.
%
%
\section{Numerical Examples and Comparison}
\label{Sect:5}
This section demonstrates the previously discussed theory for \OR{three} benchmark examples, \OR{two of which were} originally introduced in~\cite{BeexDisLatt} and~\cite{BeexHO}. The results will show that the energy balance~\eqref{E} holds along the entire loading paths for all computed solutions. In \OR{all} cases, we employ the following pair potential (in Eqs.~\eqref{Sect:5:Eq:1}~-- \eqref{Sect:5:Eq:3}, the superscripts~$\alpha\beta$ are dropped for brevity)
\begin{equation}
\phi(\widehat{r},\widehat{z}_\mathrm{p})=\frac{1}{2}\frac{EA}{r_0}(\widehat{r}-r_0-\widehat{z}_\mathrm{p})^2,
\label{Sect:5:Eq:1}
\end{equation}
i.e. the bond stiffness reads~$EA/r_0$, in accordance with the standard truss theory. Note that this definition corresponds to the rotated engineering deformation measure. The hardening potential reads as
\begin{equation}
h(\widehat{z}_\mathrm{c})=\frac{1}{\rho+1}A\sigma_0Hr_0\left(\frac{\widehat{z}_\mathrm{c}}{r_0}\right)^{\rho+1},
\label{Sect:5:Eq:2}
\end{equation}
which---by virtue of~\eqref{Sect:4:Eq:9} and~\eqref{SubSect:4.2:Eq:4}---yields the power-law hardening rule in the form
\begin{equation}
f_Y=A\sigma_0\left[1+H\left(\frac{z_\mathrm{c}}{r_0}\right)^\rho\right].
\label{Sect:5:Eq:3}
\end{equation}
Here, $f_Y$ denotes the current yield force of a bond connecting atoms~$\alpha$ and~$\beta$, whereas the initial yield force reads~$f_0=A\sigma_0$. All physical constants used throughout this section are specified in Tab.~\ref{Sect:5:Tab:1}.\footnote{Although the choice of~$h$ made above may seem academic, note that the power hardening law~\eqref{Sect:5:Eq:3} has been used by~\cite{Beex:textileReliability} to model the mechanical behaviour of woven fabrics.}
\begin{table}
	\caption{Dimensionless material and geometric parameters for \OR{all} examples.}
	\centering
	\renewcommand{\arraystretch}{1.5}
	\begin{tabular}{l|r@{}l}
	Physical parameters & \multicolumn{2}{c}{Value} \\\hline
	Young's modulus, \hfill $E$ & 1 & \\
	Cross-sectional area, \hfill $A$ & 1 & \\
	Yield stress, \hfill $\sigma_0$ & 0 & .01 \\
	Hardening modulus, \hfill $H$ & 10 & \\
	Hardening exponent, \hfill $\rho$ & 0 & .5
	\end{tabular}
	\label{Sect:5:Tab:1}
\end{table}
%
%
\subsection{Uniaxial Loading Test}
\label{SubSect:6.1}
\begin{figure}
	\centering
	\includegraphics[scale=1]{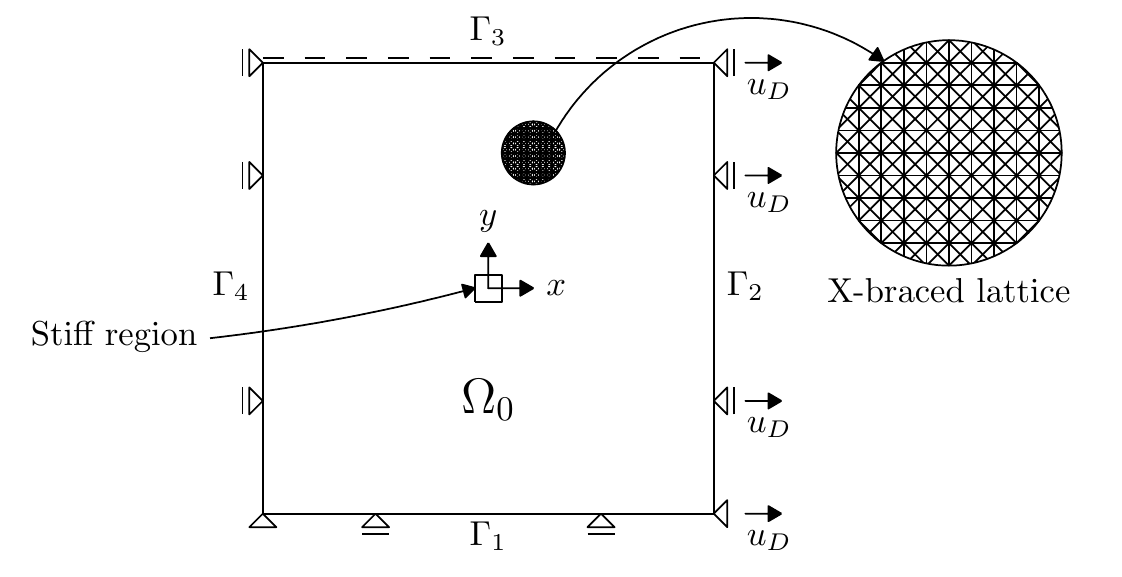}
	\caption{Scheme of the uniaxial loading test: geometry and boundary conditions.}
	\label{Sect:5:Fig:1}
\end{figure}

As a first example a uniform loading test is presented. The domain~$\Omega_0$ occupies~$100\times 100$ lattice spacings of~1 unit length; the fully resolved system consists of~$10,201$ atoms and~$40,200$ bonds. A stiff region occupying~$6\times 6$ lattice spacings in the centre represents an inhomogeneity within an otherwise homogeneous medium. The Young's modulus of the springs in the stiff domain is~$100$ times larger than elsewhere and the initial yield force~$f_0$ is infinite to prevent plastic yielding in the inclusion. All bonds on the boundary~$\partial\Omega_0 = \bigcup_{i=1}^4\Gamma_i$ (see Fig.~\ref{Sect:5:Fig:1}), have cross-sectional areas reduced to~$1/2$. The boundary conditions are set according to~\cite{BeexDisLatt}, Section~4.1:
\begin{subequations}
\label{Sect:5:Eq3}
\begin{align}
r_y(\Gamma_1) &= r_{0,y}(\Gamma_1)=-50,\nonumber\\
r_x(\Gamma_2) &= r_{0,x}(\Gamma_2)+10t=50+10t,\ t\in[0,1],\nonumber\\
r_x(\Gamma_4) &= r_{0,x}(\Gamma_4) = -50,\label{Sect:5:Eq3a}\\
\bs{r}(\Gamma_1\cap\Gamma_4)&=\bs{r}_0(\Gamma_1\cap\Gamma_4)=
\left[\begin{array}{c}
-50 \\
-50
\end{array}\right],\nonumber\\
r_y(\Gamma_3) &= r_y(\Gamma_3\cap\Gamma_4),\label{Sect:5:Eq3b}
\end{align}
\end{subequations}
where~$\bs{r}(\Gamma)$ denotes the deformed configurations of all the atoms on line segment~$\Gamma$ ($\bs{r}^\alpha=[r_x^\alpha,r_y^\alpha]^\mathsf{T}$). Time interval~$[0,1]$ is divided uniformly into~$100$ increments (i.e.~$n_T=100$ and~$T=1$, cf.~\eqref{IP}). 

\begin{figure}
	\centering
	\subfloat[S$_\mathrm{a}$]{\includegraphics[scale=0.4]{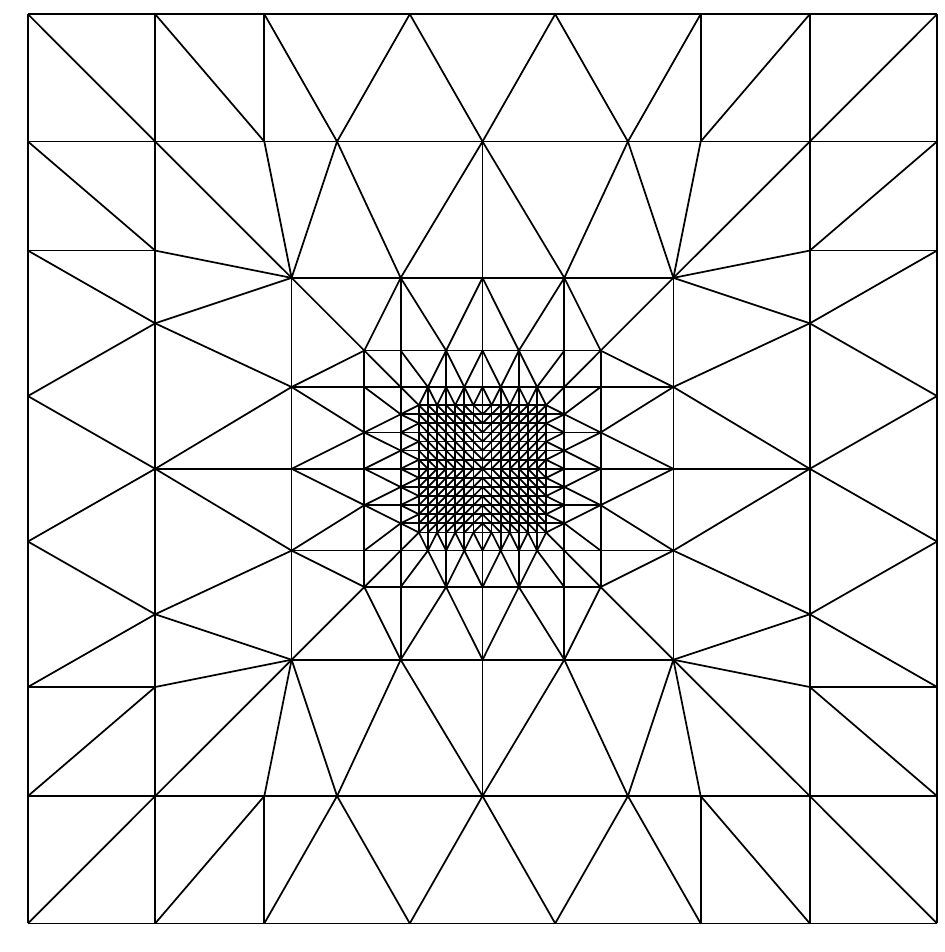}\label{Sect:5:Fig:2a}}\hspace{0.2em}
	\subfloat[S$_\mathrm{b}$]{\includegraphics[scale=0.4]{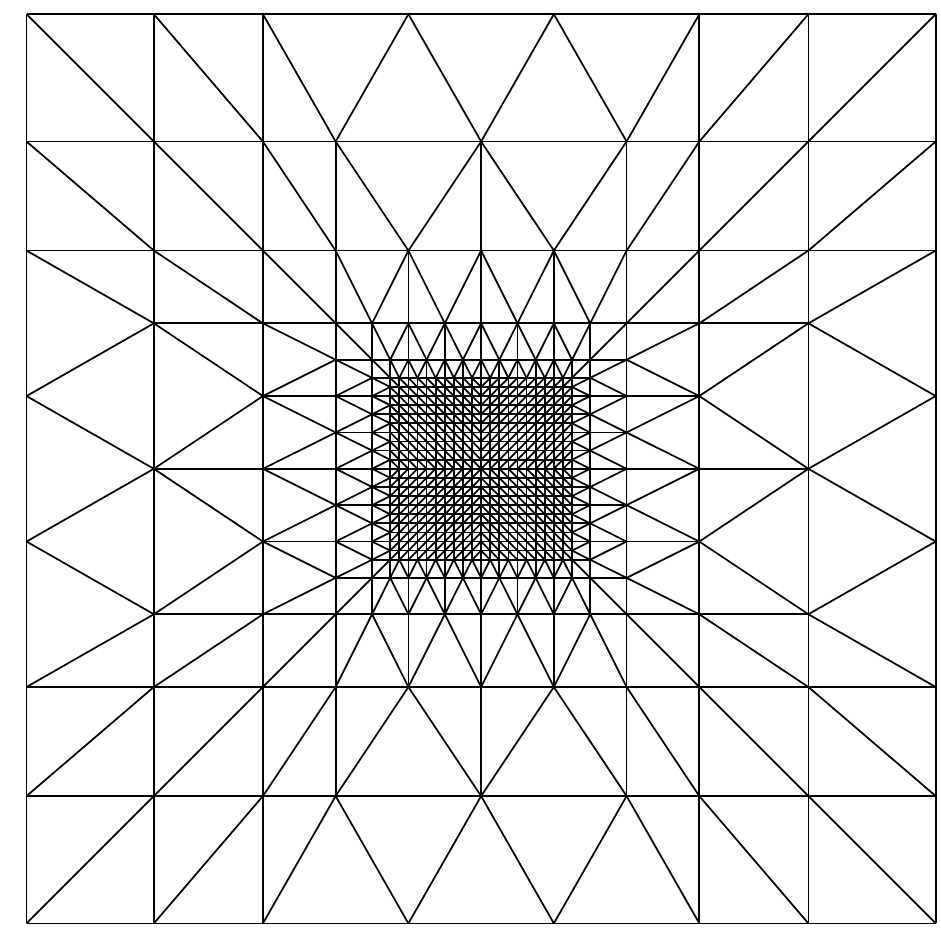}\label{Sect:5:Fig:2b}}\hspace{0.2em}
	\subfloat[S$_\mathrm{c}$]{\includegraphics[scale=0.4]{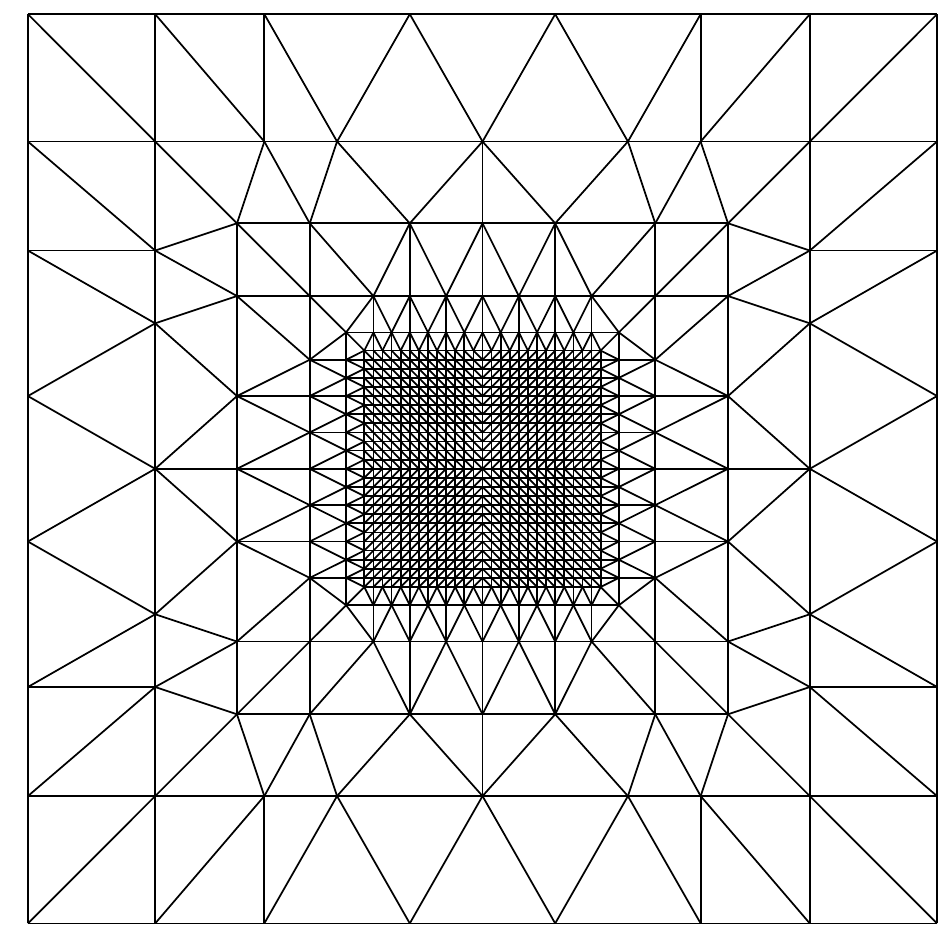}\label{Sect:5:Fig:2c}}\hspace{0.2em}
	\subfloat[S$_\mathrm{d}$]{\includegraphics[scale=0.4]{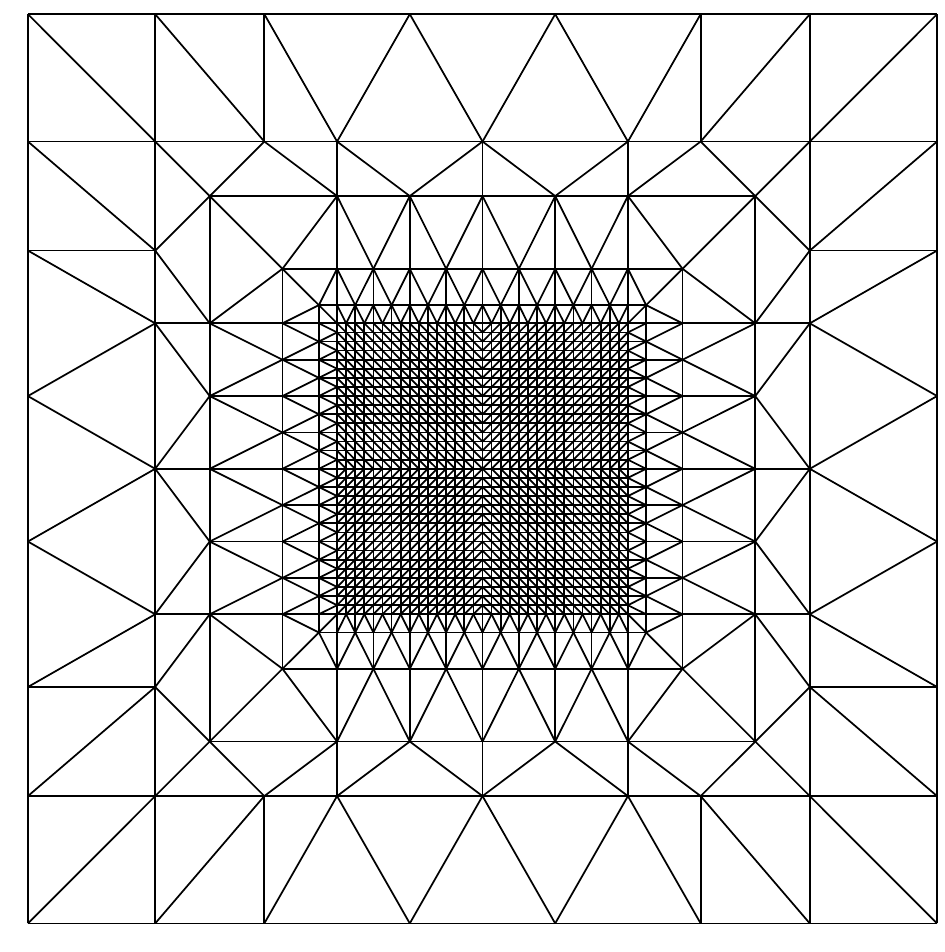}\label{Sect:5:Fig:2d}}\\
	\subfloat[U$_\mathrm{a}$]{\includegraphics[scale=0.4]{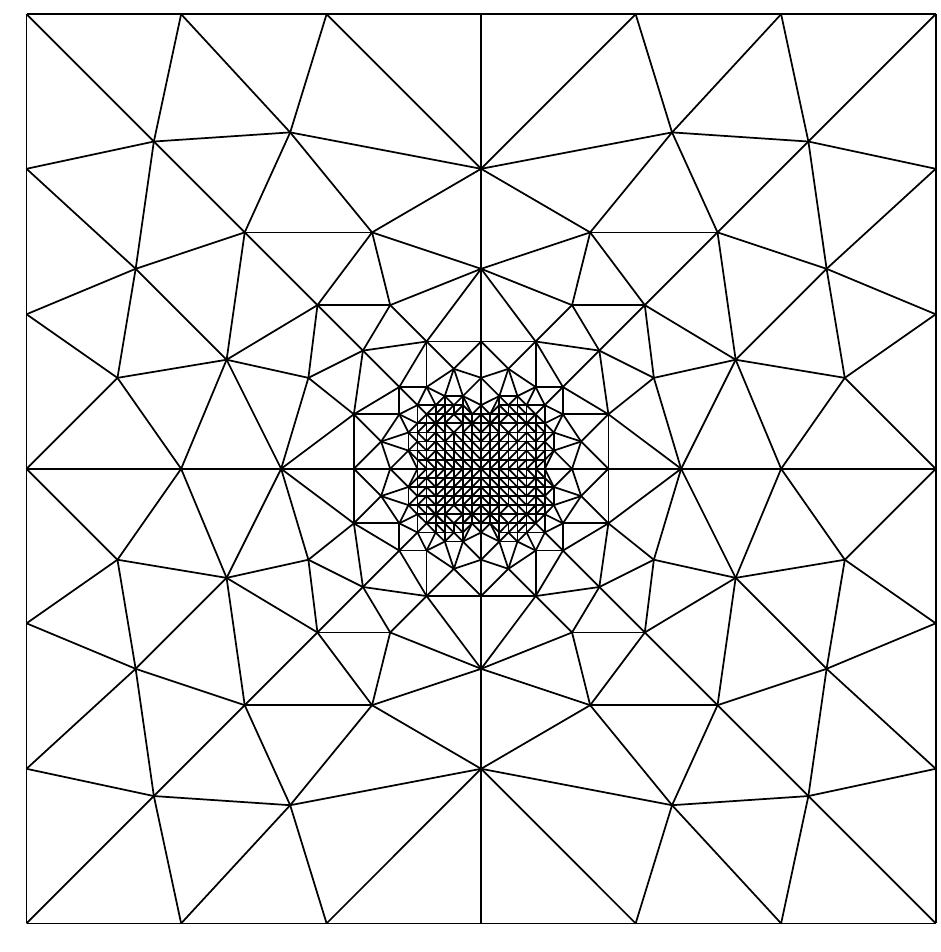}\label{Sect:5:Fig:2e}}\hspace{0.2em}
	\subfloat[U$_\mathrm{b}$]{\includegraphics[scale=0.4]{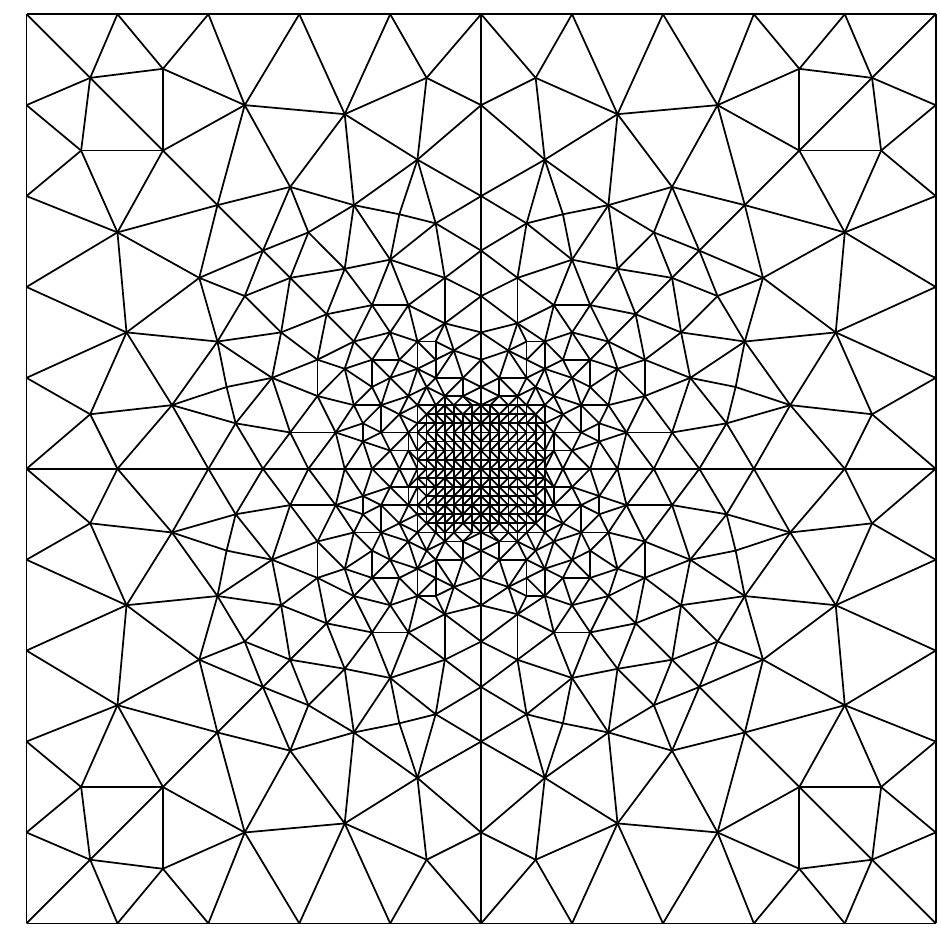}\label{Sect:5:Fig:2f}}\hspace{0.2em}
	\subfloat[U$_\mathrm{c}$]{\includegraphics[scale=0.4]{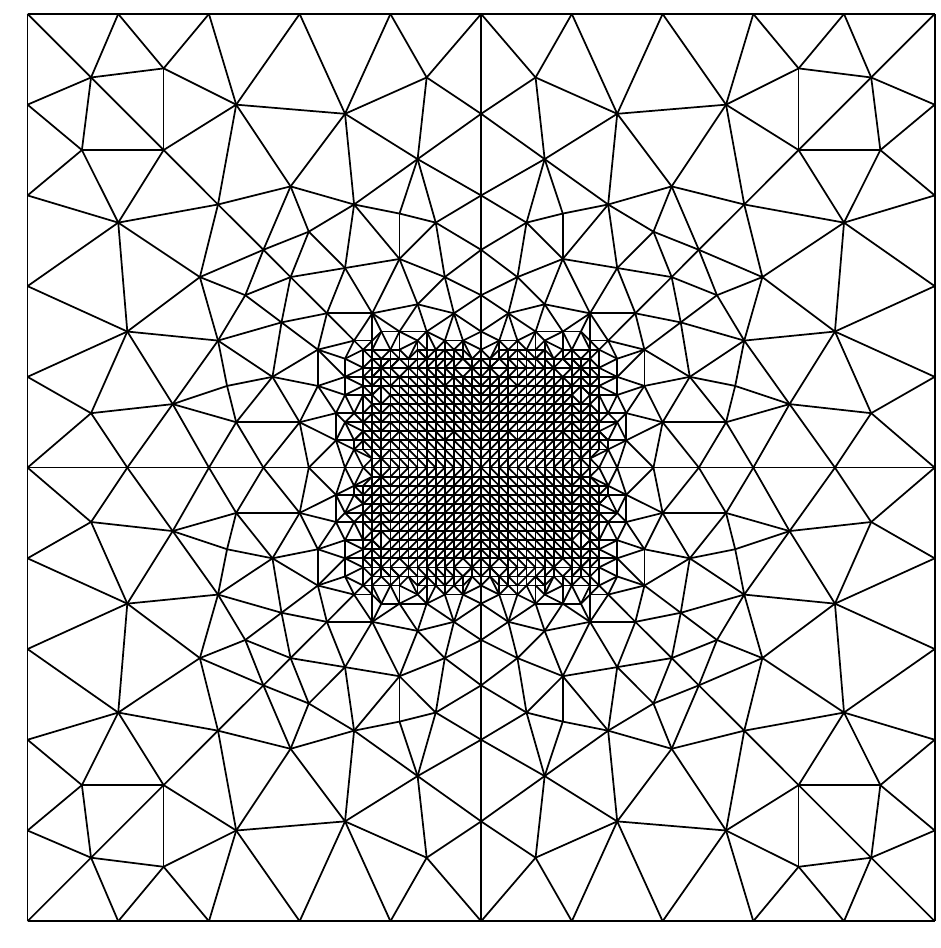}\label{Sect:5:Fig:2g}}\hspace{0.2em}
	\subfloat[U$_\mathrm{d}$]{\includegraphics[scale=0.4]{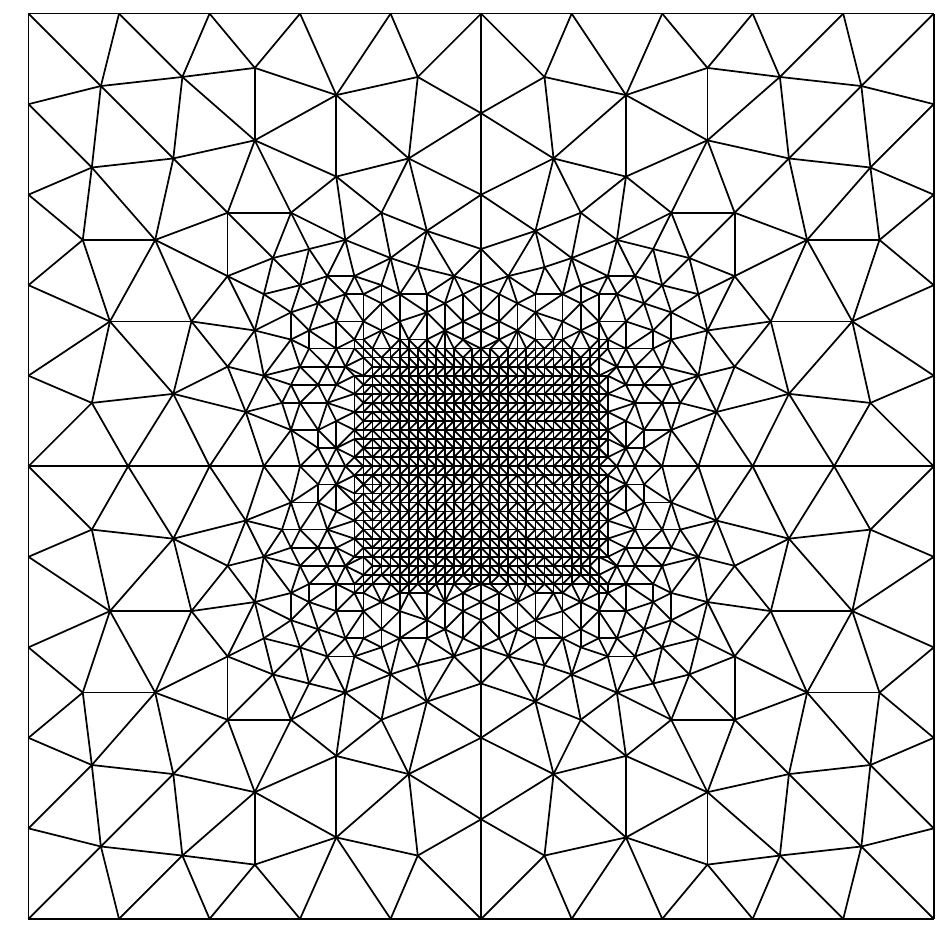}\label{Sect:5:Fig:2h}}	
	\caption{Eight triangulations for the uniaxial loading test: (a)~-- (d) structured~"S" meshes, used also in~\cite{BeexDisLatt}, and (e)~-- (h) unstructured~"U" meshes. The sizes of the fully-resolved regions, the numbers of repatoms, sampling atoms, and sampling bonds corresponding to these meshes are provided in Tab.~\ref{Sect:5:Tab:3}.}
	\label{Sect:5:Fig:2}
\end{figure}

The numerical example is performed for the exact and central summation rules, utilizing nine meshes; eight of them are depicted in Fig.~\ref{Sect:5:Fig:2} while the ninth one represents the fully-resolved lattice. In order to demonstrate the importance of the mesh quality, we use two groups of triangulations: structured~"S" (Figs.~\ref{Sect:5:Fig:2a}~-- \ref{Sect:5:Fig:2d}) and unstructured~"U" (Figs.~\ref{Sect:5:Fig:2e}~-- \ref{Sect:5:Fig:2h}), constructed such that the number of repatoms ($n_\mathrm{rep}$) is pairwise nearly the same, see Tab.~\ref{Sect:5:Tab:3}. For the unstructured meshes, the coarsening immediately starts outside the fully-refined region with a mild coarsening gradient. Consequently, the sizes of the fully-resolved regions differ.
\begin{table}
	\caption{Uniaxial loading test: the sizes of the fully-resolved regions~"size full", numbers of repatoms~$n_\mathrm{rep}$, sampling atoms~$n_\mathrm{sam}^\mathrm{ato}$, and sampling bonds~$n_\mathrm{sam}^\mathrm{bon}$ for the meshes depicted in Fig.~\ref{Sect:5:Fig:2}; "Ex" refers to the exact and "C" to the central summation rule, "S" to the structured and~"U" to the unstructured meshes.}
	\centering
	\renewcommand{\arraystretch}{1.5}
	\begin{tabular}{l@{}l|rrrrrrrr:r}
		\multicolumn{2}{c|}{Quantity} & \multicolumn{1}{c}{S$_\mathrm{a}$} & \multicolumn{1}{c}{U$_\mathrm{a}$} & \multicolumn{1}{c}{S$_\mathrm{b}$} & \multicolumn{1}{c}{U$_\mathrm{b}$} & \multicolumn{1}{c}{S$_\mathrm{c}$} & \multicolumn{1}{c}{U$_\mathrm{c}$} & \multicolumn{1}{c}{S$_\mathrm{d}$} & \multicolumn{1}{c}{U$_\mathrm{d}$} & \multicolumn{1}{:c}{full} \\\hline	
		\multicolumn{2}{c}{size full} & \multicolumn{1}{|c}{$\scriptstyle 14\times 14$} & \multicolumn{1}{c}{$\scriptstyle 8\times 8$} & \multicolumn{1}{c}{$\scriptstyle 20\times 20$} & \multicolumn{1}{c}{$\scriptstyle 12\times 12$} & \multicolumn{1}{c}{$\scriptstyle 26\times 26$} & \multicolumn{1}{c}{$\scriptstyle 20\times 20$} & \multicolumn{1}{c}{$\scriptstyle 32\times 32$} & \multicolumn{1}{c}{$\scriptstyle 24\times 24$} & \multicolumn{1}{:c}{$\scriptstyle 100\times 100$} \\
		\multicolumn{2}{c|}{$n_\mathrm{rep}$} & 349 & 337 & 597 & 537 & 893 & 969 & 1,277 & 1,113 &  10,201 \\
		\multirow{2}{*}{$n_\mathrm{sam}^\mathrm{ato}\,\bigg\{$} & Ex & 5,113 & 5,621 & 5,425 & 8,209 & 5,889 & 8,169 & 6,217 & 8,161 & \multirow{2}{*}{10,201} \\	
		& C & 597 & 593 & 929 & 1,193 & 1,245 & 1,541 & 1,697 & 1,875 &  \\
		\multirow{2}{*}{$n_\mathrm{sam}^\mathrm{bon}\,\bigg\{$} & Ex & 25,389 & 27,076 & 26,814 & 36,360 & 28,373 & 36,288 & 29,474 & 36,504 & \multirow{2}{*}{40,200} \\
		& C & 3,492 & 3,536 & 5,076 & 7,636 & 6,482 & 8,544 & 8,396 & 10,388 &  \\
	\end{tabular}
	\label{Sect:5:Tab:3}
\end{table}

Concerning the implementation of step~\eqref{AMa}, the Dirichlet boundary conditions~\eqref{Sect:5:Eq3a} are applied in the standard way. The tying condition~\eqref{Sect:5:Eq3b} is enforced by first collecting all atoms or repatoms on~$\Gamma_3$ (except for atom~$\bs{r}(\Gamma_3\cap\Gamma_4)$) in a set~$N_{\Gamma_3}$. Then, Eq.~\eqref{Sect:5:Eq3b} is rewritten as
\begin{equation}
r_y(\Gamma_3\cap\Gamma_4)-r_y^\alpha=0,\ \alpha\in N_{\Gamma_3},
\label{Sect:5:Eq4}
\end{equation}
which are globally assembled to
\begin{equation}
\bs{Cr}=\bs{0},
\label{Sect:5:Eq5}
\end{equation}
and imposed by Lagrange multipliers, i.e. by the primal-dual formulation. This yields an iterative solution of saddle-point problems (cf.~\cite{BonnansOptim}, Section~14), that reads
\begin{equation}
\left[\begin{array}{cc}
\bs{K}^i & \bs{C}^\mathsf{T}\\
\bs{C} & \bs{0}
\end{array}\right]
\left[\begin{array}{c}
\widehat{\bs{r}}^{i+1}-\widehat{\bs{r}}^i\\
\widehat{\bs{\lambda}}^{i+1}
\end{array}\right]
= 
-\left[\begin{array}{c}
\bs{f}^i \\
\bs{B}^i
\end{array}\right], \quad \OR{\bs{B}^i = \bs{C}\widehat{\bs{r}}^i},
\label{Sect:5:Eq6}
\end{equation}
for time step~$t_k$ and AM iteration~$l$, providing us with~$\bs{r}^{l+1}$ and~$\bs{\lambda}^{l+1}$ upon convergence. \OR{For QC systems, $\bs{K}^i$ is replaced with~$\bs{H}^i$ and~$\widehat{\bs{r}}^i$ with~$\widehat{\bs{r}}^i_\mathrm{rep}$.}

\begin{figure}
	\centering
	\begin{tikzpicture}
	\linespread{1}
	
	\node[inner sep=0pt] (energy) at (0,0) {
		\subfloat[energy evolutions]{\includegraphics[scale=1]{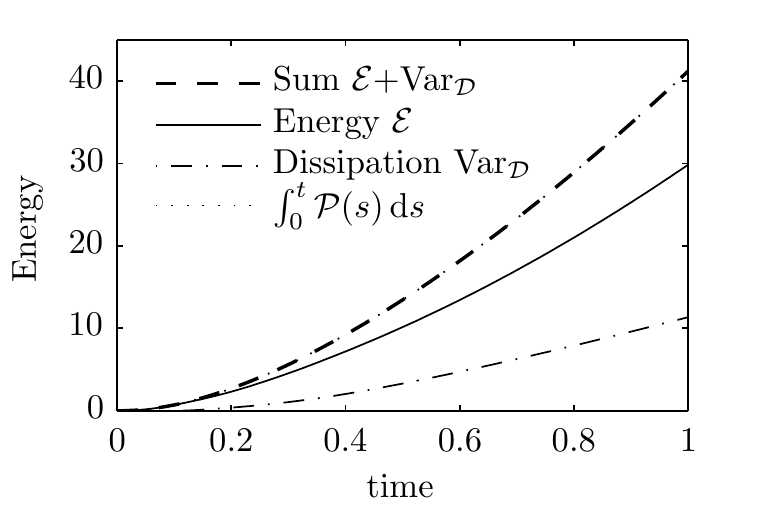}\label{Sect:5:Fig:3a}}};
	\node[inner sep=0pt] (zoom) at (8,0.325) {
		\subfloat[zoom]{\includegraphics[scale=1]{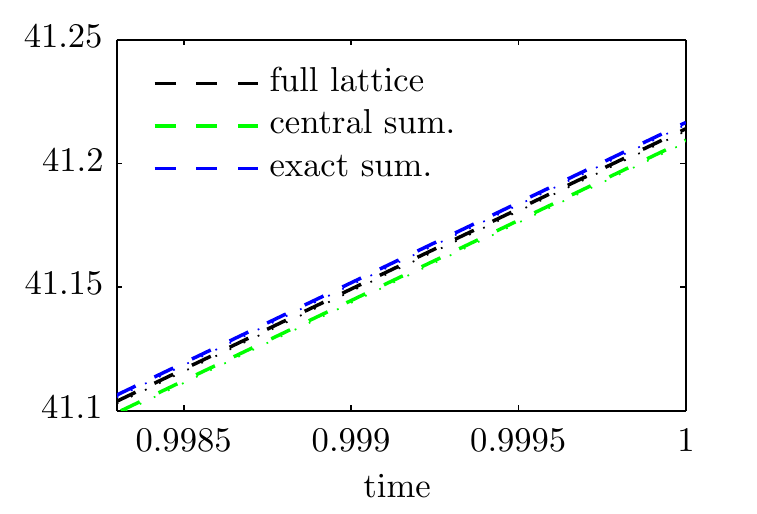}\label{Sect:5:Fig:3b}}};
	
	\draw[black, thick, dashed, rounded corners] (2.4,1.8) rectangle (3.8,2.5);
	
	\draw[black, thick, dashed] (3.1,1.8) -- (5.5,-1.13);
	\draw[black, thick, dashed] (3.1,2.5) -- (5.5,2.63);
	
	\end{tikzpicture}
	\caption{Results for the uniaxial loading test: (a)~total energy evolution paths for all meshes and both summation rules, (b)~zoom of only mesh~$\mathrm{S_a}$ (Fig.~\ref{Sect:5:Fig:2a}) for the exact and central summation rule.}
	\label{Sect:5:Fig:3}
\end{figure}
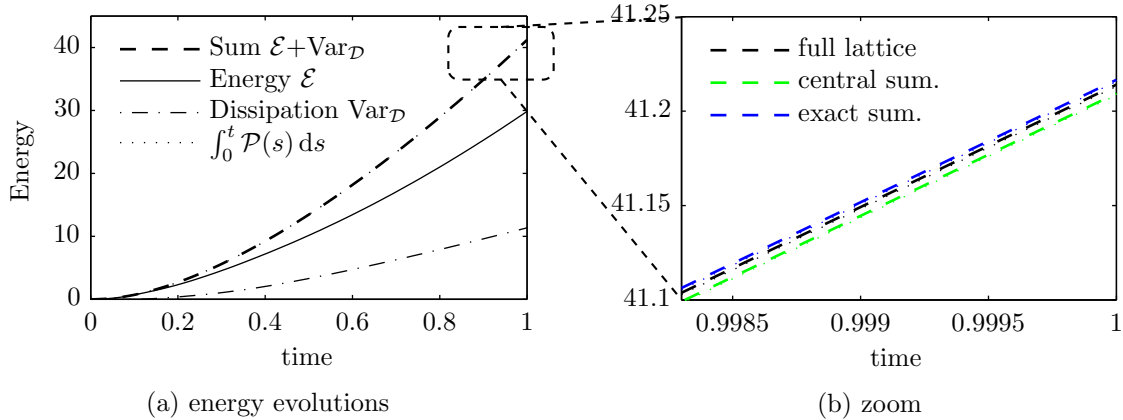

The energy profiles corresponding to the loading program~\eqref{Sect:5:Eq3} for all the meshes and both summation rules are depicted in Fig.~\ref{Sect:5:Fig:3a}. Here, we notice that the curves corresponding to the individual solutions are identical and that the errors are extremely small. Moreover, we can check that all solutions satisfy the energy balance~\eqref{E} along the entire loading path, since the thin dotted line corresponding to~$\int_{0}^{t}\mathcal{P}(s)\,\mathrm{d}s$ lies on top of the thick dashed line representing~$\mathcal{E}+\mathrm{Var}_\mathcal{D}$. Upon zooming in (Fig.~\ref{Sect:5:Fig:3b}, where only the results for the mesh~$\mathrm{S_a}$ are presented for clarity), we observe that the exact summation rule increases the system's energy with respect to the full lattice simulation, whereas the central summation rule causes the energy to be slightly lower. This behaviour is not surprising as the overall energy increases when the geometric constraints of the QC system are introduced (while using the exact summation rule). Further, since the site energies of atoms lying near the triangles' edges are higher compared to internal atoms (recall Fig.~\ref{SubSect:3.2:Fig:1}), the approximate energy for the central summation rule is slightly lower.

\begin{figure}
	\centering
	\includegraphics[scale=1]{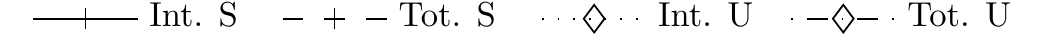}\vspace{-0.75em}\\
	\subfloat[$\varepsilon_\mathcal{E}$]{\includegraphics[scale=1]{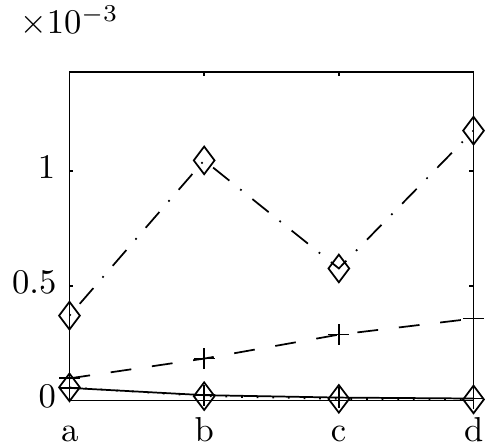}\label{Sect:5:Fig:4a}}\hspace{0.5em}
	\subfloat[$\varepsilon_{\mathrm{Var}_\mathcal{D}}$]{\includegraphics[scale=1]{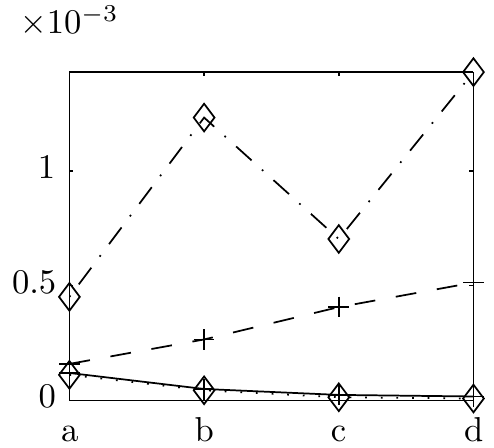}\label{Sect:5:Fig:4b}}\hspace{0.5em}
	\subfloat[$\varepsilon_{\mathcal{E}+\mathrm{Var}_\mathcal{D}}$]{\includegraphics[scale=1]{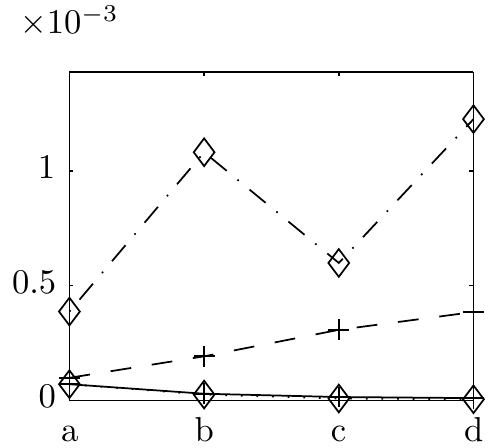}\label{Sect:5:Fig:4c}}
\caption{Results for the uniaxial loading test: relative error~$\varepsilon_{\widetilde{\Box}}$ (Eq.~\eqref{SubSect:6.1:Eq:1}) for the meshes presented in Fig.~\ref{Sect:5:Fig:2}. "Int." relates to interpolation (exact summation rule) and "Tot." to interpolation plus summation (central summation rule).}
	\label{Sect:5:Fig:4}
\end{figure}

\begin{figure}
	\centering
	\includegraphics[scale=1]{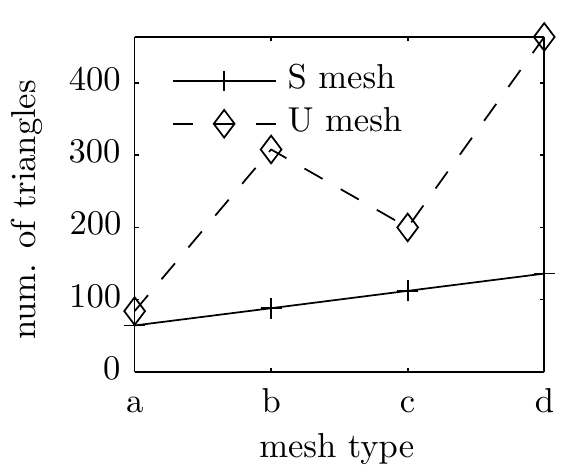}
	\caption{Results for the uniaxial loading test: number of triangles for which the central sampling atoms do not have all their neighbours within the same triangle; meshes from Fig.~\ref{Sect:5:Fig:2} used.}
	\label{Sect:5:Fig:5}
\end{figure}

Because the exact and approximate energy profiles are indistinguishable by the naked eye, we introduce a relative error measure~$L^2([0,1])$ for~$\mathcal{E}(t),\mathrm{Var}_\mathcal{D}(t)$, and~$\mathcal{E}(t)+\mathrm{Var}_\mathcal{D}(t)$. Namely,
\begin{equation}
\varepsilon_{\widetilde{\Box}} = \frac{||\Box_\mathrm{QC}-\Box_\mathrm{full}||_{L^2}}{||\Box_\mathrm{full}||_{L^2}},
\label{SubSect:6.1:Eq:1}
\end{equation}
where~$\Box$ either stands for~$\mathcal{E}(t),\mathrm{Var}_\mathcal{D}(t)$, or~$\mathcal{E}(t)+\mathrm{Var}_\mathcal{D}(t)$, and~$\widetilde{\Box}$ stands for~$\mathcal{E},\mathrm{Var}_\mathcal{D}$, or~$\mathcal{E}+\mathrm{Var}_\mathcal{D}$. The subscript~"QC" denotes results obtained from the QC simulations, whereas the subscript~"full" represents the results computed for the fully-resolved system. The error measure~$\varepsilon_{\widetilde{\Box}}$ is presented for the various meshes in Fig.~\ref{Sect:5:Fig:4}. We further distinguish between the error due to interpolation (denoted as "Int.") and the total error due to interpolation plus summation (denoted as "Tot."). Thus, "Int." relates to the error obtained for the exact summation rule, whereas "Tot." relates to the error obtained for the central summation rule. The largest value amounts to~$1.43 \times 10^{-3}$ and corresponds to the dissipation~$\mathrm{Var}_\mathcal{D}$, cf. Fig.~\ref{Sect:5:Fig:4b}. Whilst the error due to interpolation behaves in a reasonable way, i.e. decreases with an increasing dimension of the projection basis, the interpolation plus summation error surprisingly behaves in the opposite way (see Fig.~\ref{Sect:5:Fig:4}), i.e. increases with an increasing number of sampling atoms. Since the total error is dominated by summation, this behaviour can be related to the mesh topology. Namely, to the number of triangles for which the central sampling atom and all its neighbours are not contained in the same triangle.\footnote{Note that in accordance with the QC convention, a triangular element is considered as a closed set. Consequently, the atoms lying on element's edges or vertices are contained in that triangle.} Such a conjecture is supported by the perfect match between the total error profiles in Fig.~\ref{Sect:5:Fig:4c} and the mesh characteristics in Fig.~\ref{Sect:5:Fig:5}. Numerically quantified, the Pearson's correlation coefficient for these data amounts to~$0.998$ (structured mesh) and~$0.963$ (unstructured mesh). This result supports the idea of a direct transition between the fully-resolved and interpolated regions, or motivates the use of rapid mesh coarsening rather than a mild one. On the other hand, the overall error is so small that this behaviour can also be deemed singular. Finally, let us note that while the energy errors (global error measures) are negligible, the relative errors of the internal variables (local error measures) are still noticeable, cf.~\cite{BeexDisLatt}, Section~4.3. But this is consistent with one's expectations as the QC aims to closely approximate the global incremental energy~$\Pi^k$, recall Section~\ref{SubSect:3.2}.

Concerning the mesh types, we conclude from Fig.~\ref{Sect:5:Fig:4} that the performance in terms of the interpolation error is nearly the same, though the unstructured meshes are slightly more accurate. In terms of the total error, however, the structured meshes perform evidently better. The structured ones are also more efficient, as for them the corresponding numbers of sampling atoms ($n_\mathrm{sam}^{\mathrm{ato}}$) and sampling bonds ($n_\mathrm{sam}^{\mathrm{bon}}$) are systematically lower, see Tab.~\ref{Sect:5:Tab:3}. Recall that the numbers of repatoms ($n_\mathrm{rep}$) remain comparable. Such behaviour can again be related to the mesh topology and the mesh coarsening gradient. Combined results from Fig.~\ref{Sect:5:Fig:4} and Tab.~\ref{Sect:5:Tab:3} reveal that by accepting energy errors up to~$2\,\%$, the number of degrees of freedom reduces up to the factor of~$30$. In the case of sampling atoms, the reduction is up to the factor of~$17$.

The highly accurate energy reconstruction of the QC method for the uniform loading test is attributed to two aspects. First, the (plastic) deformation field is more or less piecewise constant over the deformed domain~$\Omega(t)$. Hence, it is well captured by constant approximations within triangles. Second, perturbations near the inhomogeneity are resolved accurately by all employed meshes. Consequently, the summation error outweighs the interpolation error. These two reasons and the example presented in~\cite{BeexHO}, Sections~4.2 and~5.2 motivate the following test, in which the deformation field due to bending is linear rather than constant and more pronounced interpolation errors are expected.
%
%
\subsection{Pure Bending Test}
\label{SubSect:6.2}
\begin{figure}
	\centering
	\subfloat[geometry]{\includegraphics[scale=1]{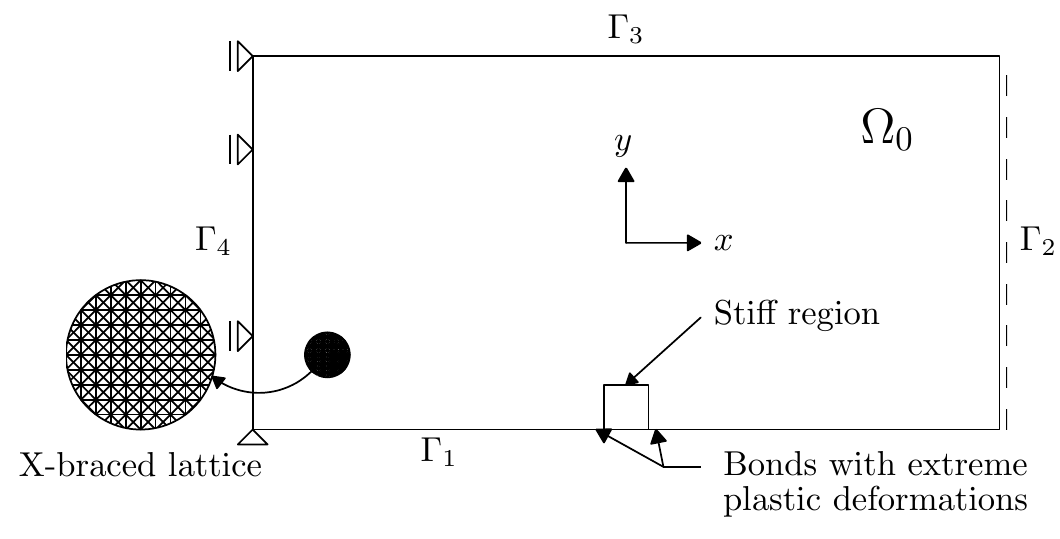}\label{Sect:5:Fig:6a}}\hspace{1em}
	\subfloat[deformed state]{\includegraphics[scale=1]{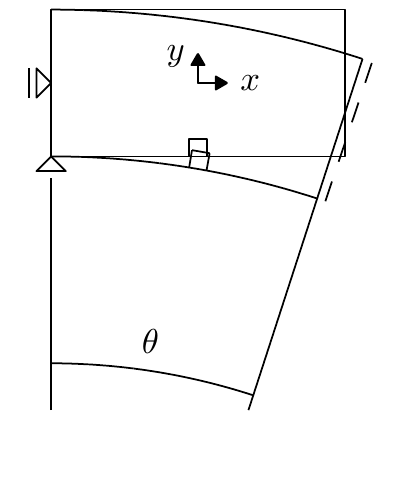}\label{Sect:5:Fig:6b}}	
	\caption{Scheme of the pure bending test: geometry, boundary conditions, and deformed state.}
	\label{Sect:5:Fig:6}
\end{figure}

In this example, the domain~$\Omega_0$ is exposed to pure bending around the $z$-axis, cf. Fig.~\ref{Sect:5:Fig:6}. An inhomogeneity of~$6\times 6$ lattice spacings is situated at the bottom edge, in which the trusses have a $100$-times higher Young's modulus and an infinite initial yield force~$f_0$ prevents plastic yielding. The entire domain occupies~$200\times 100$ lattice units and comprises~$20,301$ atoms connected by~$80,300$ bonds. The boundary conditions for~$\partial\Omega_0 = \bigcup_{i=1}^4\Gamma_i$ read as follows
\begin{subequations}
	\label{SubSect:6.2:Eq1}
	\begin{align}
	r_x(\Gamma_4) &= r_{0,x}(\Gamma_4)=-100,\nonumber\\
	\bs{r}(\Gamma_1\cap\Gamma_4)&=\bs{r}_0(\Gamma_1\cap\Gamma_4)=
	\left[\begin{array}{c}
	-100 \\ 
	-50
	\end{array}\right],\label{SubSect:6.2:Eq1a}\\
	\frac{r_x^{\alpha_{j+1}}-r_x^{\alpha_j}}{r_y^{\alpha_{j+1}}-r_y^{\alpha_j}} &= \tan\theta,\ j=1,\dots,n_{\Gamma_2}-1,\label{SubSect:6.2:Eq1b}
	\end{align}
\end{subequations}
where we have collected~$n_{\Gamma_2}$ atoms or repatoms on~$\Gamma_2$ in the set~$N_{\Gamma_2}$; note that~$\alpha_j = N_{\Gamma_2}(j)$ denotes the $j$-th element of~$N_{\Gamma_2}$. All atoms in~$N_{\Gamma_2}$ are sorted from bottom to top, i.e.~$r_y^{\alpha_{j+1}} > r_y^{\alpha_j}$, and hence, $\bs{r}^{\alpha_1} = \bs{r}(\Gamma_1\cap\Gamma_2)$ and~$\bs{r}^{\alpha_{n_{\Gamma_2}}}=\bs{r}(\Gamma_2\cap\Gamma_3)$. Condition~\eqref{SubSect:6.2:Eq1b} therefore requires that~$\bs{r}(\Gamma_2)$ follows a straight line that is allowed to freely move in space and has a slope~$\theta$, cf. Fig.~\ref{Sect:5:Fig:6b}. Note that atoms can slide frictionlessly along this line but cannot move in the perpendicular direction. Equation~\eqref{SubSect:6.2:Eq1b} can then be rewritten as
\begin{equation}
r_x^{\alpha_{j+1}}-r_x^{\alpha_j}+(r_y^{\alpha_j}-r_y^{\alpha_{j+1}})\tan\theta=0,\ j=1,\dots,n_{\Gamma_2}-1,
\label{SubSect:6.2:Eq2}
\end{equation}
and globally assembled as
\begin{equation}
\bs{C}(\theta)\bs{r}=\bs{0},
\label{SubSect:6.2:Eq3}
\end{equation}
in analogy to Eqs.~\eqref{Sect:5:Eq4} and~\eqref{Sect:5:Eq5}. Since the constraints now depend on~$\theta(t)$ through Eq.~\eqref{SubSect:6.2:Eq4} below, matrix~$\bs{C}$ changes for each time step~$t_k$. To accomodate this, $\bs{C}$ is replaced by~$\bs{C}_k=\bs{C}(\theta(t_k))$ in Eq.~\eqref{Sect:5:Eq6}. The overall deformation process is finally parametrized as
\begin{equation}
\theta(t)=\frac{\pi}{6}\,t,\ t\in [0,1],
\label{SubSect:6.2:Eq4}
\end{equation}
and the time interval is divided into~100 uniform increments (i.e.~$n_T=100$, $T=1$).

\begin{figure}
	\centering
	\subfloat[$\mathrm{B_a}$]{\includegraphics[scale=0.6]{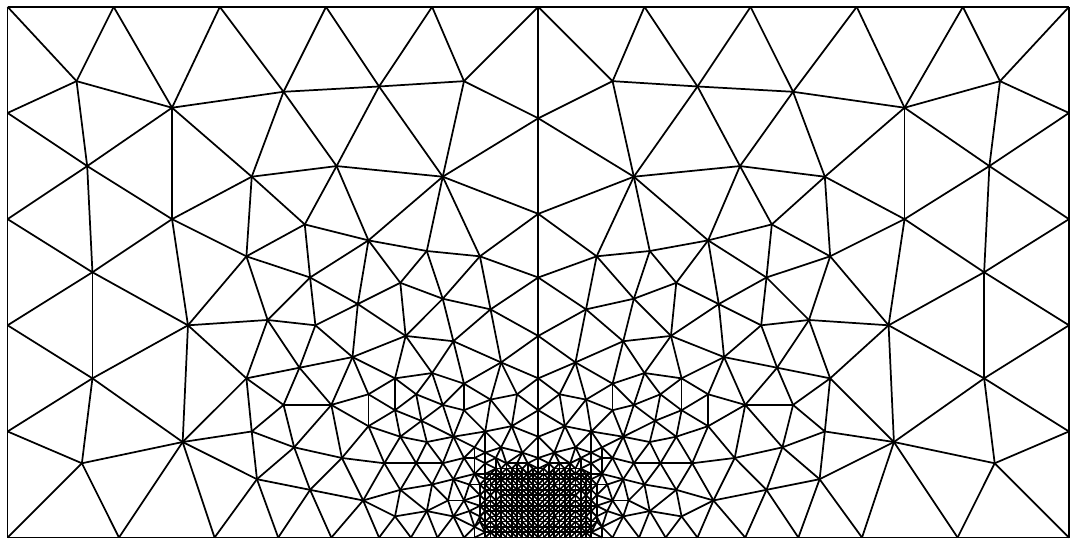}\label{SubSect:5.2:Fig:2a}}\hspace{0.2em}
	\subfloat[$\mathrm{B_b}$]{\includegraphics[scale=0.6]{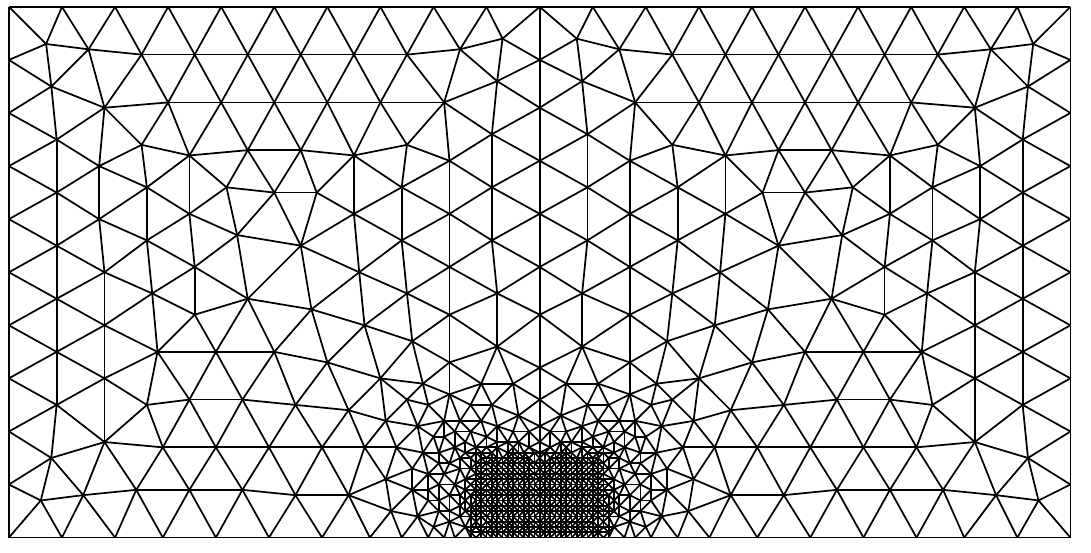}\label{SubSect:5.2:Fig:2c}}\\
	\subfloat[$\mathrm{B_c}$]{\includegraphics[scale=0.6]{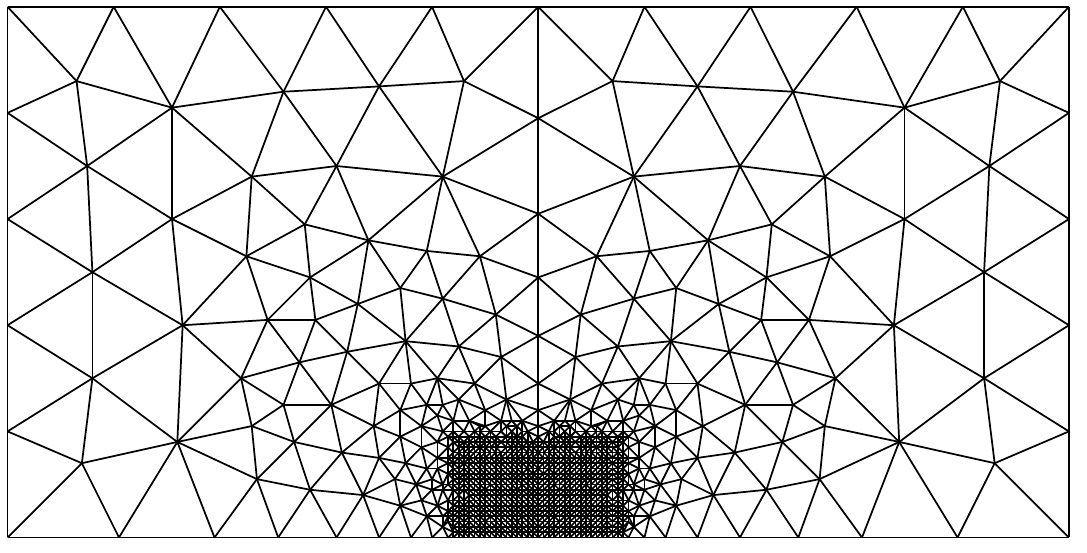}\label{SubSect:5.2:Fig:2d}}\hspace{0.2em}
	\subfloat[$\mathrm{B_d}$]{\includegraphics[scale=0.6]{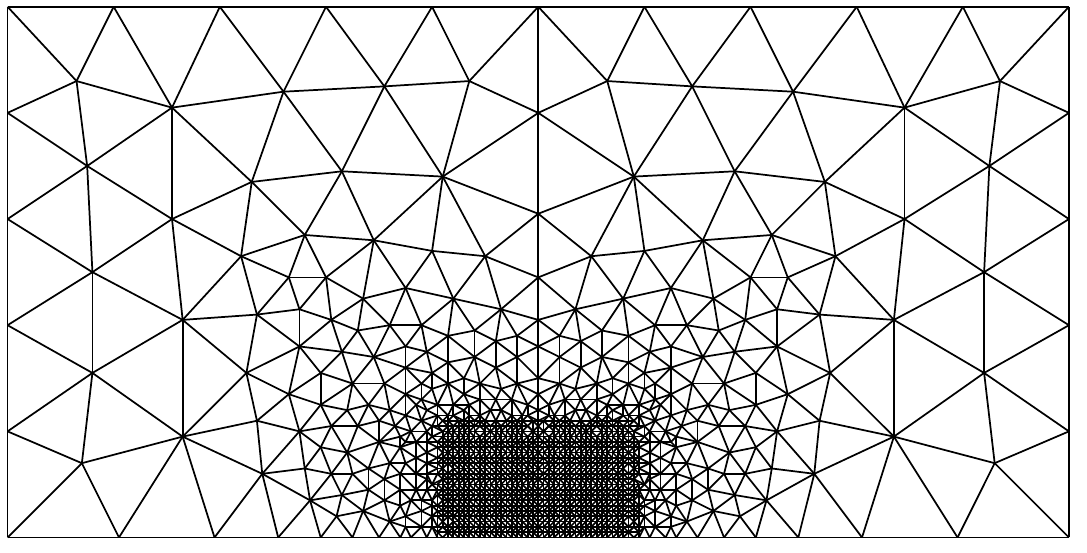}\label{SubSect:5.2:Fig:2e}}
	\caption{The four triangulations for the pure bending test; the sizes of the fully-resolved regions, the numbers of repatoms, sampling atoms, and sampling bonds are presented in Tab.~\ref{Sect:5:Tab:4}.}
	\label{SubSect:5.2:Fig:2}
\end{figure}

The numerical example is again studied for the two summation rules, but now only for five unstructured meshes. Four of them are shown in Fig.~\ref{SubSect:5.2:Fig:2}, the fifth one represents the full lattice. The corresponding numbers of atoms, repatoms, and sampling bonds can be found in Tab.~\ref{Sect:5:Tab:4}.
\begin{table}
	\caption{Data for the pure bending test: the sizes of the fully-resolved regions~"size full", numbers of repatoms~$n_\mathrm{rep}$, sampling atoms~$n_\mathrm{sam}^\mathrm{ato}$, and sampling bonds~$n_\mathrm{sam}^\mathrm{bon}$ for the meshes depicted in Fig.~\ref{SubSect:5.2:Fig:2}; "Ex" refers to the exact and "C" to the central summation rule.}
	\centering
	\renewcommand{\arraystretch}{1.5}
	\begin{tabular}{l@{}l|rrrr:r}
		\multicolumn{2}{c|}{Quantity} & \multicolumn{1}{c}{$\mathrm{B_a}$} & \multicolumn{1}{c}{$\mathrm{B_b}$} & \multicolumn{1}{c}{$\mathrm{B_c}$} & \multicolumn{1}{c}{$\mathrm{B_d}$} & \multicolumn{1}{:c}{full} \\\hline
		\multicolumn{2}{c|}{size full} & \multicolumn{1}{|c}{$\scriptstyle 14\times 10$} & \multicolumn{1}{c}{$\scriptstyle 20\times 13$} & \multicolumn{1}{c}{$\scriptstyle 26\times 16$} & \multicolumn{1}{c}{$\scriptstyle 32\times 19$} & \multicolumn{1}{:c}{$\scriptstyle 200\times 100$} \\
		\multicolumn{2}{c|}{$n_\mathrm{rep}$} & 507 & 812 & 894 & 1,194 & 20,301 \\
		\multirow{2}{*}{$n_\mathrm{sam}^\mathrm{ato}\,\bigg\{$} & Ex & 10,709 & 13,989 & 10,693 & 11,301 & \multirow{2}{*}{20,301} \\	
		& C & 941 & 1,558 & 1,342 & 1,872 & \\
		\multirow{2}{*}{$n_\mathrm{sam}^\mathrm{bon}\,\bigg\{$} & Ex & 51,094 & 66,990 & 51,136 & 52,432 & \multirow{2}{*}{80,300} \\
		& C & 5,789 & 9,574 & 7,215 & 10,120 & \\
	\end{tabular}
	\label{Sect:5:Tab:4}
\end{table}

The energy evolution paths for all meshes are shown in Fig.~\ref{SubSect:5.2:Fig:3} for the exact summation rule, and in Fig.~\ref{SubSect:5.2:Fig:4} for the central summation rule. Here, the differences between the solutions are more pronounced compared to the previous example, though they are still quite small. The results show that the energy equality~\eqref{E} holds again. Upon closer inspection, we notice that the interpolation error (Fig.~\ref{SubSect:5.2:Fig:3b}) is smallest for mesh~$\mathrm{B_b}$, which has the smallest overall element size. So the error is dominated by the limitation of constant strain triangles to capture bending, and is almost insensitive to the size of the fully-resolved region. Instead of refining the triangulation in the coarse part of the domain, an alternative approach for decreasing the interpolation error would be to use higher-order approximations, as shown e.g. in~\cite{BeexHO}. Furthermore, the total error due to interpolation and summation (Fig.~\ref{SubSect:5.2:Fig:4b}) is smaller than the interpolation error alone, meaning that the two errors partially compensate. 

\ignore{In this example, the total error is driven by interpolation rather than summation in contrast to the uniform loading test.}
\begin{figure}
	\centering
	\begin{tikzpicture}
	\linespread{1}
	
	\node[inner sep=0pt] (energy) at (0,0) {
		\subfloat[exact summation rule]{\includegraphics[scale=1]{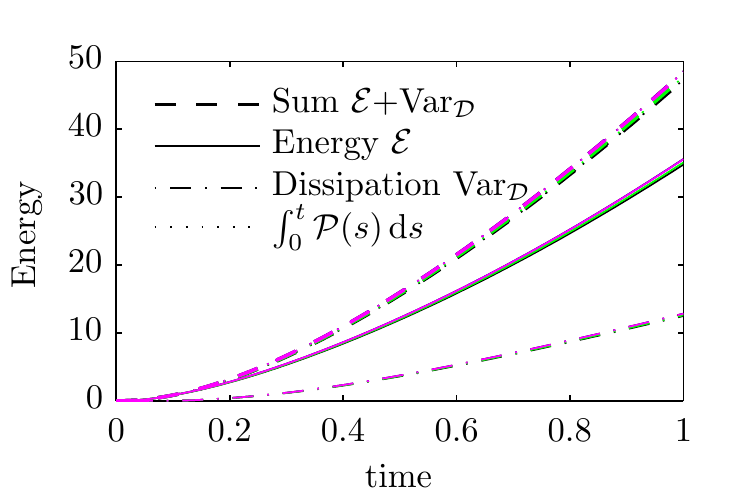}\label{SubSect:5.2:Fig:3a}}};
	\node[inner sep=0pt] (zoom) at (8,0.325) {
		\subfloat[zoom]{\includegraphics[scale=1]{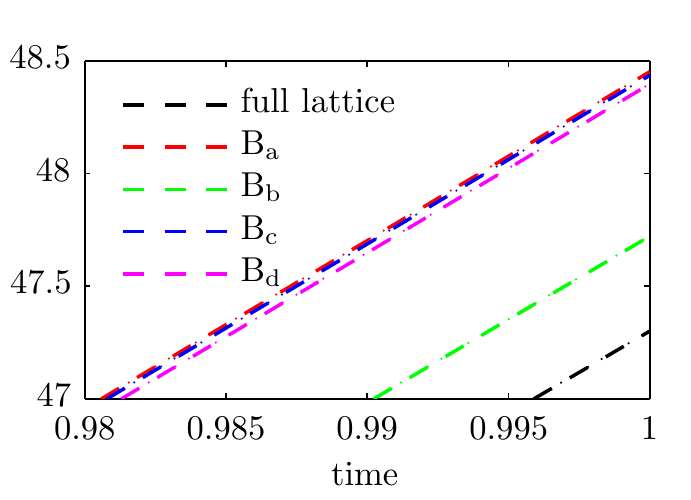}\label{SubSect:5.2:Fig:3b}}};
	
	\draw[black, thick, dashed, rounded corners] (2.2,1.5) rectangle (3.6,2.2);
	
	\draw[black, thick, dashed] (2.9,1.5) -- (5.35,-1.08);
	\draw[black, thick, dashed] (2.9,2.2) -- (5.35,2.35);
	
	\end{tikzpicture}
	\caption{Results for the pure bending test: (a)~total energy evolution paths, (b)~zoom; the different meshes from Fig.~\ref{SubSect:5.2:Fig:2} using the exact summation rule are compared to the full-lattice solution.}
	\label{SubSect:5.2:Fig:3}
\end{figure}
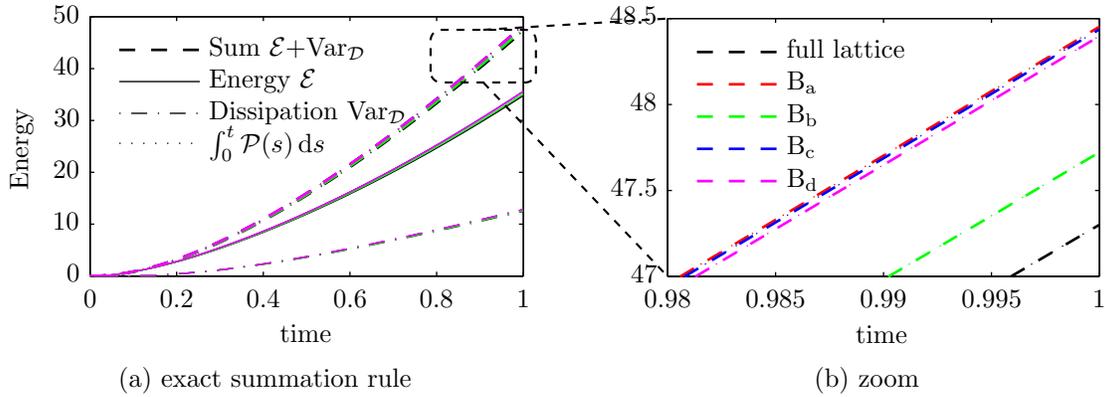
\begin{figure}
	\centering
	\begin{tikzpicture}
	\linespread{1}
	
	\node[inner sep=0pt] (energy) at (0,0) {
		\subfloat[central summation rule]{\includegraphics[scale=1]{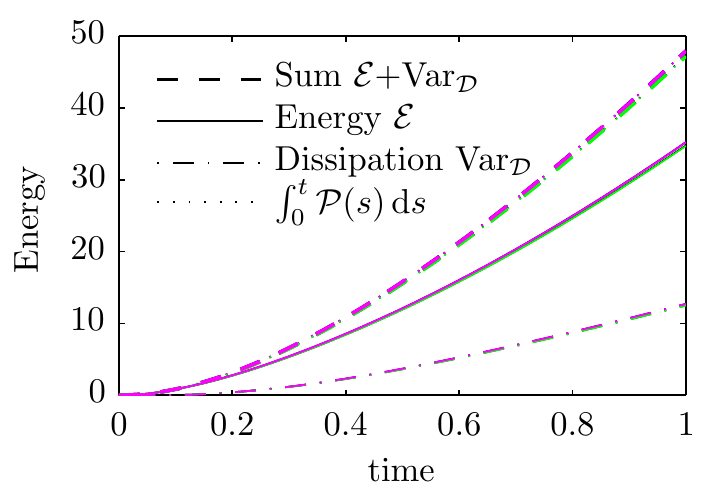}\label{SubSect:5.2:Fig:4a}}};
	\node[inner sep=0pt] (zoom) at (8,0.2) {
		\subfloat[zoom]{\includegraphics[scale=1]{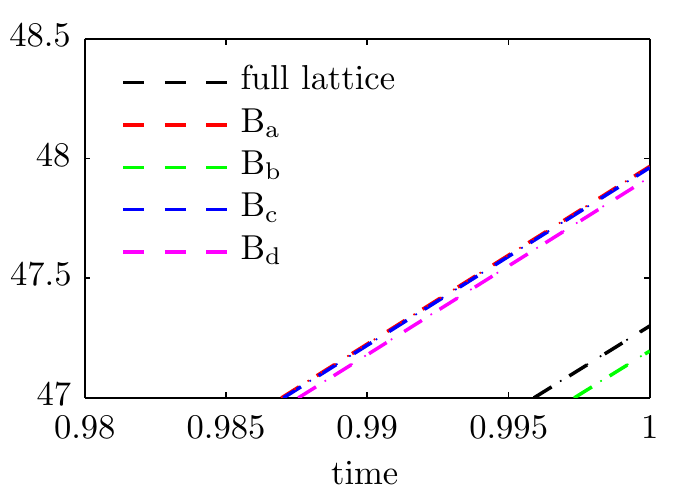}\label{SubSect:5.2:Fig:4b}}};
	
	\draw[black, thick, dashed, rounded corners] (2.1,1.6) rectangle (3.5,2.3);
	
	\draw[black, thick, dashed] (2.8,1.6) -- (5.35,-1.2);
	\draw[black, thick, dashed] (2.8,2.3) -- (5.35,2.44);
	
	\end{tikzpicture}
	\caption{Results for the pure bending test: (a)~energy evolution paths, (b)~zoom; the different meshes from Fig.~\ref{SubSect:5.2:Fig:2} using the central summation rule are compared to the full-lattice solution.}
	\label{SubSect:5.2:Fig:4}
\end{figure}
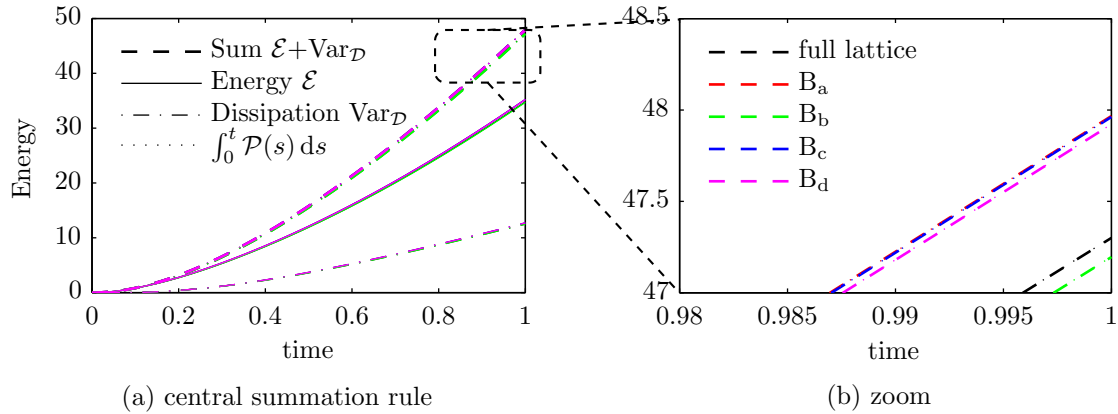

\begin{figure}
	\centering
	\subfloat[$\varepsilon_\mathcal{E}$]{\includegraphics[scale=1]{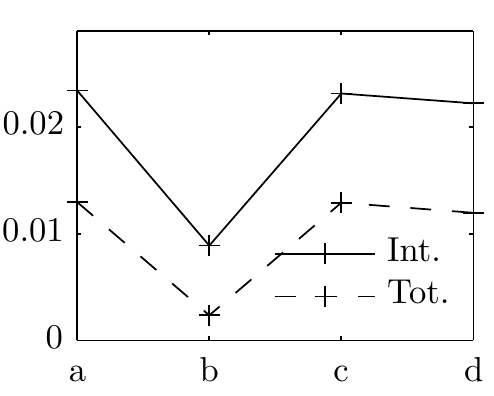}\label{SubSect:5.2:Fig:5a}}\hspace{0.5em}
	\subfloat[$\varepsilon_{\mathrm{Var}_\mathcal{D}}$]{\includegraphics[scale=1]{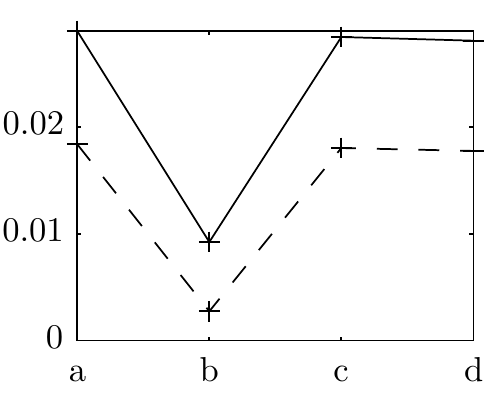}\label{SubSect:5.2:Fig:5b}}\hspace{0.5em}
	\subfloat[$\varepsilon_{\mathcal{E}+\mathrm{Var}_\mathcal{D}}$]{\includegraphics[scale=1]{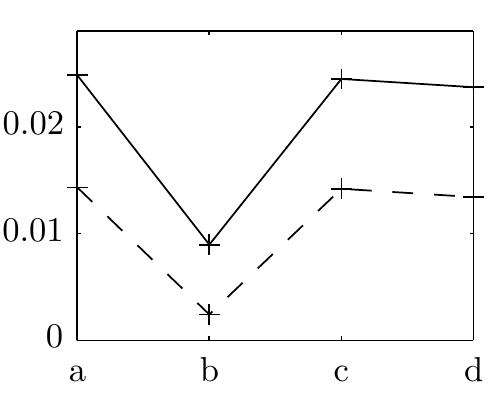}\label{SubSect:5.2:Fig:5c}}
	\caption{Results for the pure bending test: relative errors~$\varepsilon_{\widetilde{\Box}}$ defined in Eq.~\eqref{SubSect:6.1:Eq:1} for the various meshes depicted in Fig.~\ref{SubSect:5.2:Fig:2}. "Int." relates to interpolation (exact summation rule) and "Tot." to interpolation plus summation (central summation rule).}
	\label{SubSect:5.2:Fig:5}
\end{figure}

The relative error~$\varepsilon_{\widetilde{\Box}}$, defined in Eq.~\eqref{SubSect:6.1:Eq:1}, can be found in Fig.~\ref{SubSect:5.2:Fig:5} for the four meshes and the two summation rules. The best agreement achieved for mesh~$\mathrm{B_b}$ is due to the finer mesh resolution over the entire domain~$\Omega$; the largest error amounts to~$28.98 \times 10^{-3}$ attained for mesh~$\mathrm{B_a}$, for dissipation~$\mathrm{Var}_\mathcal{D}$ presented in Fig.~\ref{SubSect:5.2:Fig:5b}. Comparing the increase in the number of repatoms with respect to error, we conclude from Fig.~\ref{SubSect:5.2:Fig:5} and Tab.~\ref{Sect:5:Tab:4} that by accepting an energy error below~$3\,\%$, we gain a reduction in the number of degrees of freedom up to the factor of~$40$, and in the number of sampling atoms up to the factor of~$22$. The dependency of the error in Fig.~\ref{SubSect:5.2:Fig:5} on the mesh topology is more complicated than in the previous example. Specifically, the Pearson's correlation coefficient between the number of triangles with central sampling atoms that do not have all the neighbours in the same triangle and the "Tot." error in Fig.~\ref{SubSect:5.2:Fig:5c} drops to~$-0.051$, indicating no dependence. 

\begin{figure}
	\centering
	\includegraphics[scale=1]{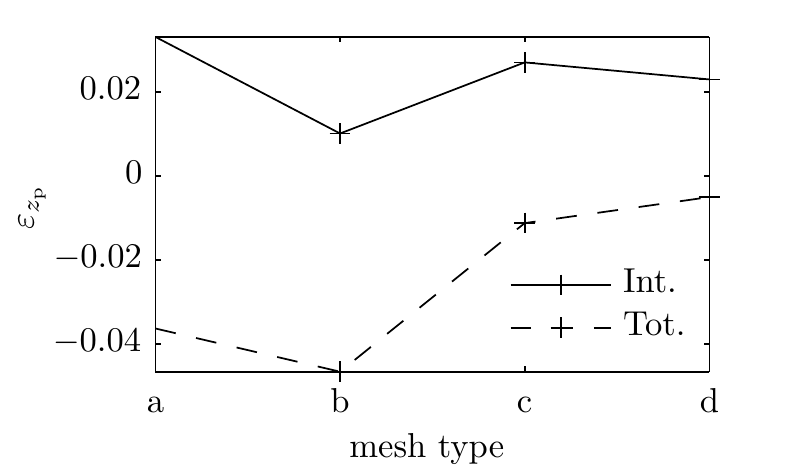}\label{SubSect:5.2:Fig:6a}
	\caption{Results for the pure bending test: relative error~$\varepsilon_{z_\mathrm{p}}(t = 1)$ defined in Eq.~\eqref{SubSect:6.2:Eq8} for various meshes depicted in Fig.~\ref{SubSect:5.2:Fig:2}. "Int." relates to interpolation (exact summation rule) and "Tot." to interpolation plus summation (central summation rule).}
	\label{SubSect:5.2:Fig:6}
\end{figure}

Finally, the following quantity is examined
\begin{equation}
\varepsilon_{z_\mathrm{p}}(t) = \frac{||\bs{z}_{\mathrm{p},\mathrm{sam}}^\mathrm{QC}(t)||_{\infty}-||\bs{z}_{\mathrm{p}}^\mathrm{full}(t)||_{\infty}}{||\bs{z}_{\mathrm{p}}^\mathrm{full}(t)||_{\infty}},
\label{SubSect:6.2:Eq8}
\end{equation}
measuring the relative error between full-lattice~"full" and approximate~"QC" solutions for extreme plastic deformations attained at~$\Gamma_1$ alongside the inhomogeneity, cf. Fig.~\ref{Sect:5:Fig:6a}. Recall that for some~$\bs{z} \in \mathbb{R}^n$, the norm in Eq.~\eqref{SubSect:6.2:Eq8} reads as~$||\bs{z}||_\infty = \max\{|z_1|,\dots,|z_n|\}$. Quantity~$\varepsilon_{z_\mathrm{p}}(t = 1)$ is presented for the different meshes and summation rules in Fig.~\ref{SubSect:5.2:Fig:6}. It can be seen that the size of the fully resolved region has again less influence on the error than the mesh refinement in the coarse domain. Because the interpolation increases the energy of the system and the summation slightly underestimates it, the two effects again cancel each other to some extent. Notice that this does not hold for mesh~$\mathrm{B_b}$, for which the summation error dominates.
%
%
\subsection{\OR{Indentation Test}}
\label{SubSect:6.3}
\OR{
In the final example, we consider a homogeneous X-braced lattice occupying a rectangular domain~$\Omega_0$ of~$256 \times 128$ lattice spacings, containing~$33,153$ atoms and~$131,456$ bonds, cf. Fig.~\ref{SubSect:6.3:Fig:1}. In accordance with the previous examples, the lattice spacings in both directions are of unit length. The boundary conditions are prescribed as
\begin{equation}
\bs{r}(\Gamma_1 \cup \Gamma_2 \cup \Gamma_4) = \bs{r}_0(\Gamma_1 \cup \Gamma_2 \cup \Gamma_4),
\label{SubSect:6.3:Eq1}
\end{equation}
and the remaining part of the boundary, $\Gamma_3$, is left free for potential frictionless contact with a rigid circular indenter. The indenter is characterized by its radius, $r_\mathrm{I} = 32$, and centre
\begin{equation}
\bs{c}_\mathrm{I}(t) = \left[
\begin{array}{c}
0 \\
r_{y,0}(\Gamma_3) + r_\mathrm{I}(1 - t/4)
\end{array}
\right], \quad t \in [0, 1].
\label{SubSect:6.3:Eq2}
\end{equation}
The time interval is again uniformly divided into~$100$ increments, i.e.~$n_T=100$ and~$T=1$. The inequality constraints associated with contact between the lattice and the indenter read
\begin{equation}
g_j(\bs{r}^{\alpha_j}) = r_\mathrm{I}^2 - || \bs{r}^{\alpha_j} - \bs{c}_\mathrm{I}||_2^2 \leq 0, \quad j = 1, \dots, n_{\Gamma_3},
\label{SubSect:6.3:Eq3}
\end{equation}
where we have collected the~$n_{\Gamma_3}$ atoms or repatoms lying on~$\Gamma_3$ part of the boundary in the set~$N_{\Gamma_3}$; recall that~$\alpha_j = N_{\Gamma_3}(j)$ is the $j$-th element of~$N_{\Gamma_3}$. These constraints are incorporated through the primal-dual formulation as before, \cite{BonnansOptim}. To that purpose, we iteratively form a set of active constraints
\begin{equation}
\mathcal{A} = \big\{ j \in \{1,\dots,n_{\Gamma_3}\} \, | \, g_j(\bs{r}^{\alpha_j}) \geq 0 \mbox{ and associated Lagrange multiplier } \lambda_j \geq 0 \big\}.
\label{SubSect:6.3:Eq5}
\end{equation}
For inactive constraints, i.e. for~$j \in\{1,\dots,n_{\Gamma_3}\}\backslash \mathcal{A}$, the associated~$\lambda_j = 0$. For all active constraints~$j \in \mathcal{A}$, the inequality conditions~\eqref{SubSect:6.3:Eq3} are enforced as equality constraints
\begin{equation}
\bs{g}(\bs{r}) = \bs{0}
\label{SubSect:6.3:Eq6}
\end{equation}
and implemented using Eq.~\eqref{Sect:5:Eq6}, where we substitute
\begin{equation}
\begin{aligned}
\bs{K}^i &= \bs{K}(\widehat{\bs{r}}^i) = \left.\frac{\partial^2 \Pi_\mathrm{red}^k}{\partial\widehat{\bs{r}}\partial\widehat{\bs{r}}}\right|_{\widehat{\bs{r}} = \widehat{\bs{r}}^i} + \left.\sum_{j \in \mathcal{A}} \widehat{\lambda}_j\frac{\partial^2 g_j(\widehat{\bs{r}}^{\alpha_j})}{\partial\widehat{\bs{r}}\partial\widehat{\bs{r}}}\right|_{\widehat{\bs{r}} = \widehat{\bs{r}}^i,\widehat{\bs{\lambda}} = \widehat{\bs{\lambda}}^i},
\\
\bs{C}^i &= \bs{C}(\widehat{\bs{r}}^i) = \left.\left(\frac{\partial\bs{g}(\widehat{\bs{r}})}{\partial\widehat{\bs{r}}}\right)^\mathsf{T}\right|_{\widehat{\bs{r}} = \widehat{\bs{r}}^i}, 
\quad
\bs{B}^i = \bs{g}(\widehat{\bs{r}}^i).
\end{aligned}
\label{SubSect:6.3:Eq7}
\end{equation}
Because the tangent~$\bs{C}(\widehat{\bs{r}}^i)$ now depends on~$\widehat{\bs{r}}^i$, it must be updated every Newton iteration, in contrast to the previous two examples. Possible violations of the inequality constraints are checked after convergence of Newton's algorithm. If a change in~$\mathcal{A}$ occurs, Newton's algorithm is called again and the entire procedure is repeated until convergence of~$\mathcal{A}$. The same strategy applies also to QC systems.

The numerical example is studied for the exact and central summation rule and again for five meshes. Four of them are depicted in Fig.~\ref{SubSect:6.3:Fig:2}, whereas the fifth represents the full lattice. Two meshes are regular and two irregular, pairwise with nearly the same number of repatoms and identical sizes of the fully-resolved regions, cf. Tab.~\ref{Sect:5:Tab:5}.

\begin{figure}
	\centering
	\includegraphics[scale=1]{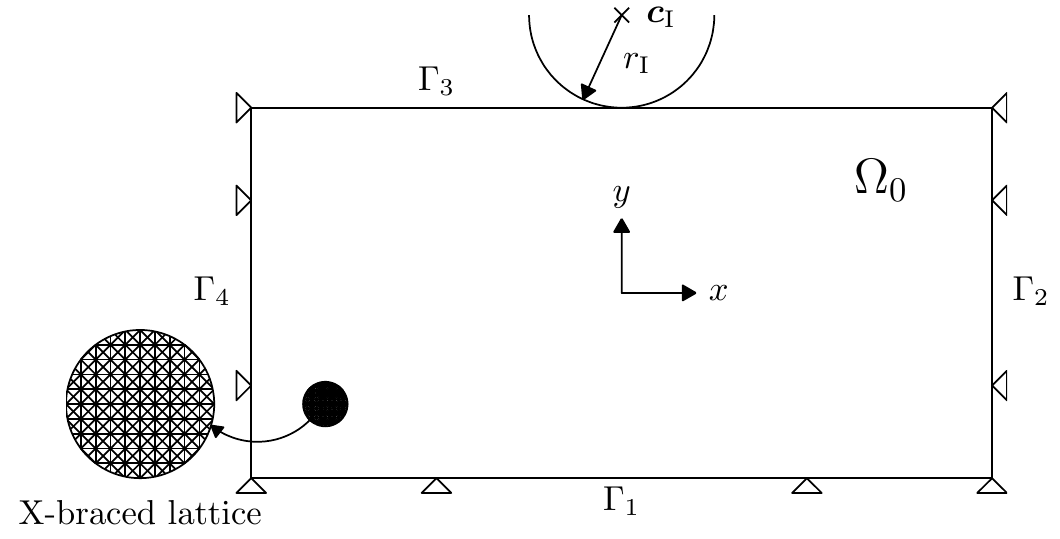}
	\caption{Scheme of the indentation test: geometry and boundary conditions.}
	\label{SubSect:6.3:Fig:1}
\end{figure}
\begin{figure}
	\centering
	\subfloat[$\mathrm{I_a}$]{\includegraphics[scale=0.6]{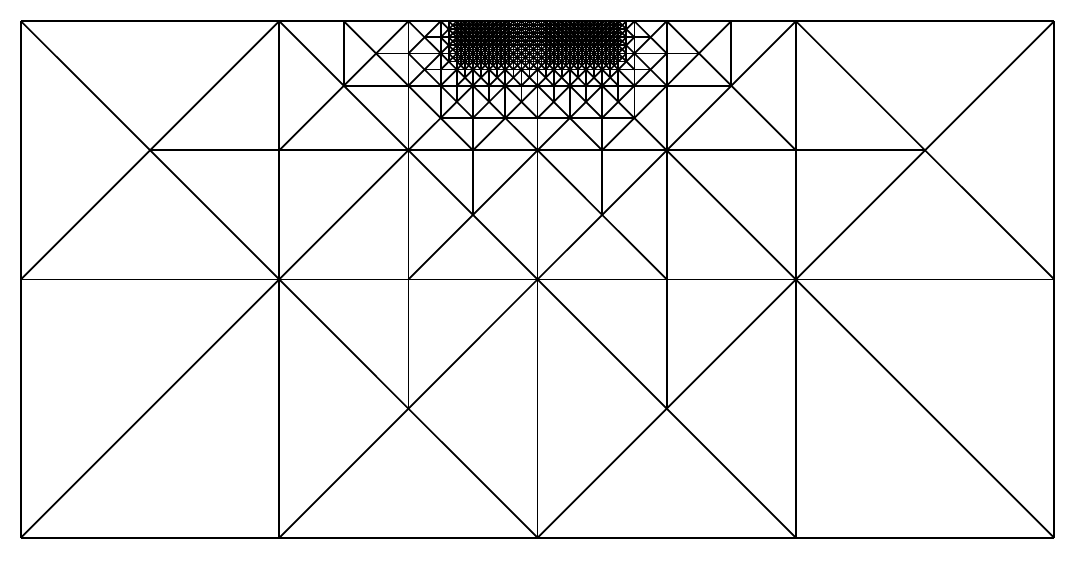}\label{SubSect:6.3:Fig:2a}}\hspace{0.2em}
	\subfloat[$\mathrm{I_b}$]{\includegraphics[scale=0.6]{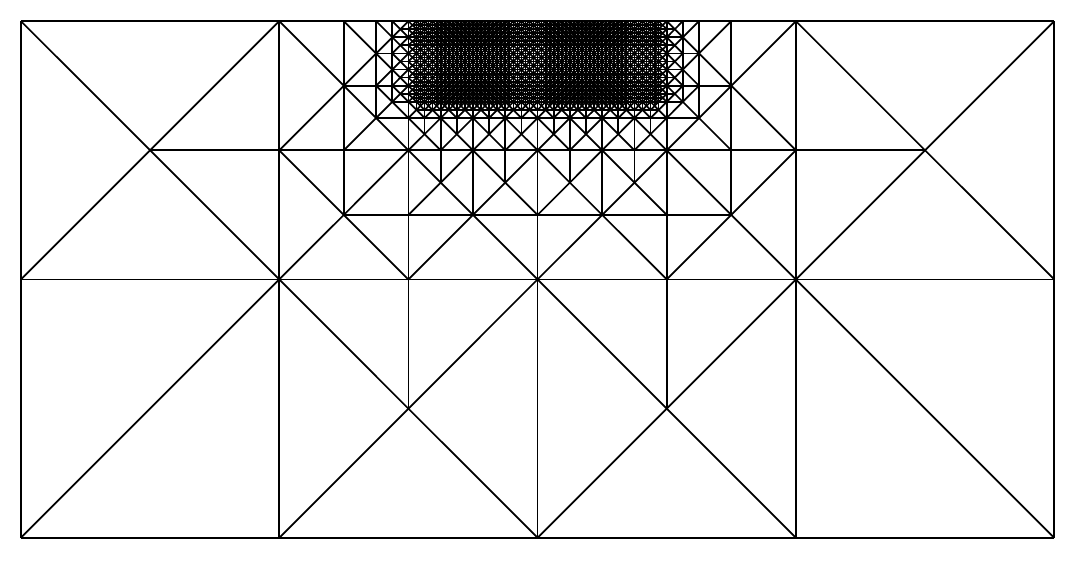}\label{SubSect:6.3:Fig:2c}}\\
	\subfloat[$\mathrm{I_c}$]{\includegraphics[scale=0.6]{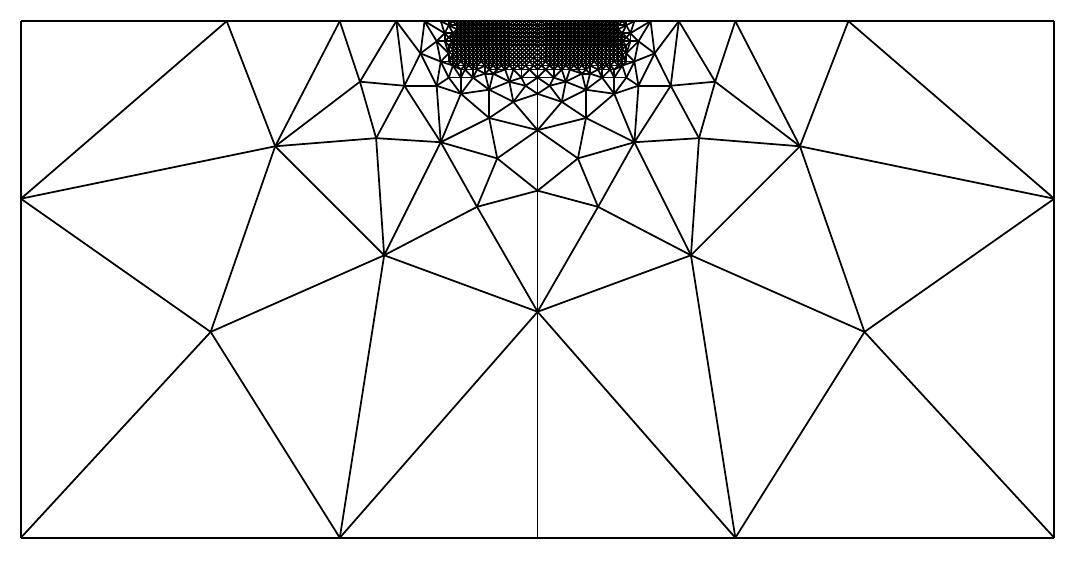}\label{SubSect:6.3:Fig:2d}}\hspace{0.2em}
	\subfloat[$\mathrm{I_d}$]{\includegraphics[scale=0.6]{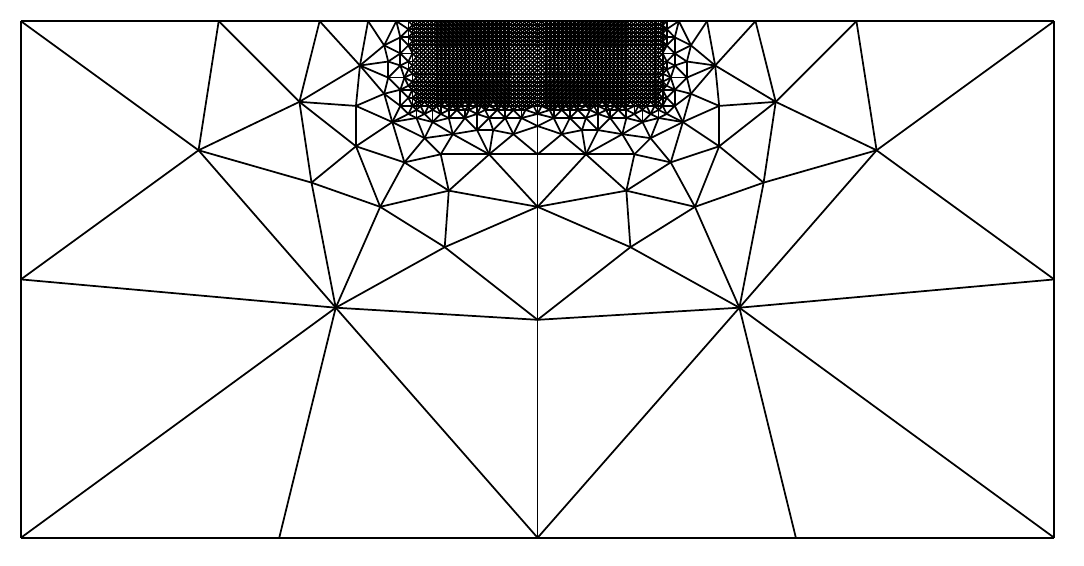}\label{SubSect:6.3:Fig:2e}}
	\caption{The four triangulations for the indentation test; the sizes of the fully-resolved regions, the numbers of repatoms, sampling atoms, and sampling bonds are presented in Tab.~\ref{Sect:5:Tab:5}.}
	\label{SubSect:6.3:Fig:2}
\end{figure}
\begin{table}
	\caption{Data for the indentation test: the sizes of the fully-resolved regions~"size full", the numbers of repatoms~$n_\mathrm{rep}$, sampling atoms~$n_\mathrm{sam}^\mathrm{ato}$, and sampling bonds~$n_\mathrm{sam}^\mathrm{bon}$ for the meshes depicted in Fig.~\ref{SubSect:6.3:Fig:2}; "Ex" refers to the exact and "C" to the central summation rule.}
	\centering
	\renewcommand{\arraystretch}{1.5}
	\begin{tabular}{l@{}l|rrrr:r}
		\multicolumn{2}{c|}{Quantity} & \multicolumn{1}{c}{$\mathrm{I_a}$} & \multicolumn{1}{c}{$\mathrm{I_b}$} & \multicolumn{1}{c}{$\mathrm{I_c}$} & \multicolumn{1}{c}{$\mathrm{I_d}$} & \multicolumn{1}{:c}{full} \\\hline
		\multicolumn{2}{c|}{size full} & \multicolumn{1}{|c}{$\scriptstyle 40\times 10$} & \multicolumn{1}{c}{$\scriptstyle 60\times 20$} & \multicolumn{1}{c}{$\scriptstyle 40\times 10$} & \multicolumn{1}{c}{$\scriptstyle 60\times 20$} & \multicolumn{1}{:c}{$\scriptstyle 256\times 128$} \\
		\multicolumn{2}{c|}{$n_\mathrm{rep}$} & 603 & 1,487 & 617 & 1,510 & 33,153 \\
		\multirow{2}{*}{$n_\mathrm{sam}^\mathrm{ato}\,\bigg\{$} & Ex & 6,249 & 7,383 & 7,531 & 8,771 & \multirow{2}{*}{33,153} \\	
		& C & 876 & 1,857 & 818 & 1,807 & \\
		\multirow{2}{*}{$n_\mathrm{sam}^\mathrm{bon}\,\bigg\{$} & Ex & 32,066 & 36,746 & 36,388 & 41,224 & \multirow{2}{*}{131,456} \\
		& C & 4,300 & 8,352 & 3,922 & 8,153 & \\
	\end{tabular}
	\label{Sect:5:Tab:5}
\end{table}

A detailed view of the deformed configuration at~$t = 1$ for the full lattice solution can be found in Fig.~\ref{SubSect:6.3:Fig:3a}. Note that only one half is shown thanks to symmetry. The dissipation localizes below the indenter through a plastic shear band, indicated in red. The reaction force of the indenter is presented in Fig.~\ref{SubSect:6.3:Fig:3b}, where we notice a good agreement between the QC simulations with the full lattice solution, especially for structured meshes~$\mathrm{I}_\mathrm{a}$ and~$\mathrm{I}_\mathrm{b}$. Because the differences between the solid and dashed lines are negligible, we can conclude that the choice of the summation rule does not influence the results and the errors are dominated by interpolation, similarly to the pure bending test in Section~\ref{SubSect:6.2}.
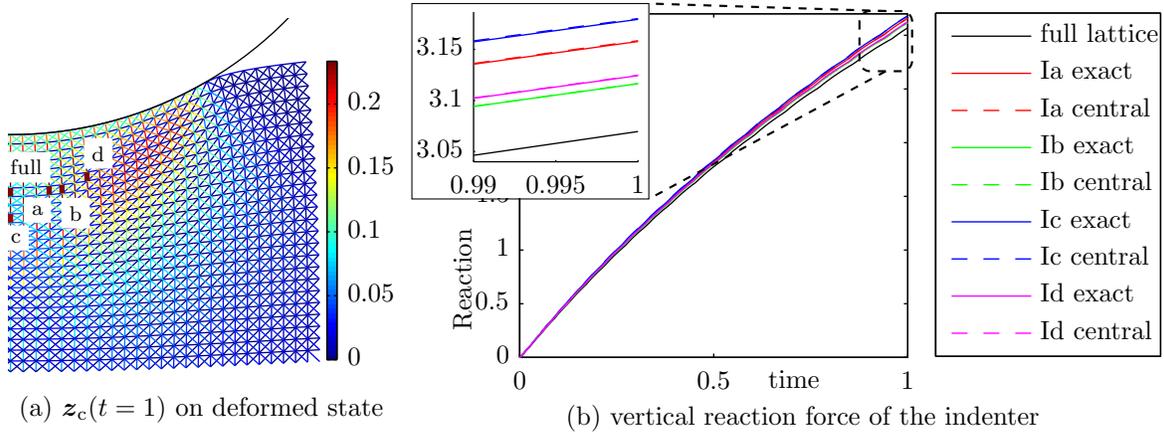
\begin{figure}
	\centering
	\begin{tikzpicture}
	\linespread{1}
	
		\node[inner sep=0pt] (Bcrack) at (1,0) {
			\subfloat[$\bs{z}_\mathrm{c}(t = 1)$ on deformed state]{\includegraphics[scale=1]{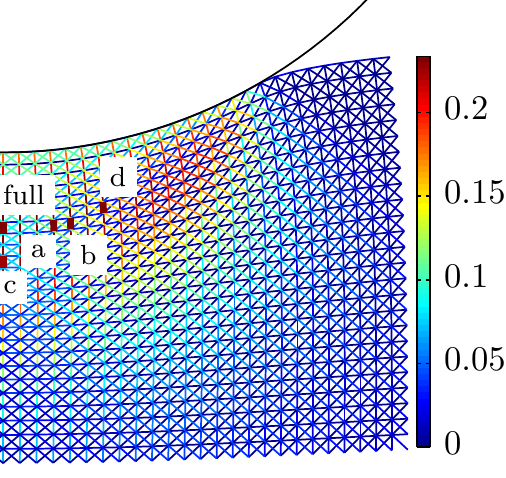}\label{SubSect:6.3:Fig:3a}}
		};
		\node[inner sep=0pt] (Breact) at (9,0.2) {
			\subfloat[vertical reaction force of the indenter]{\includegraphics[scale=1]{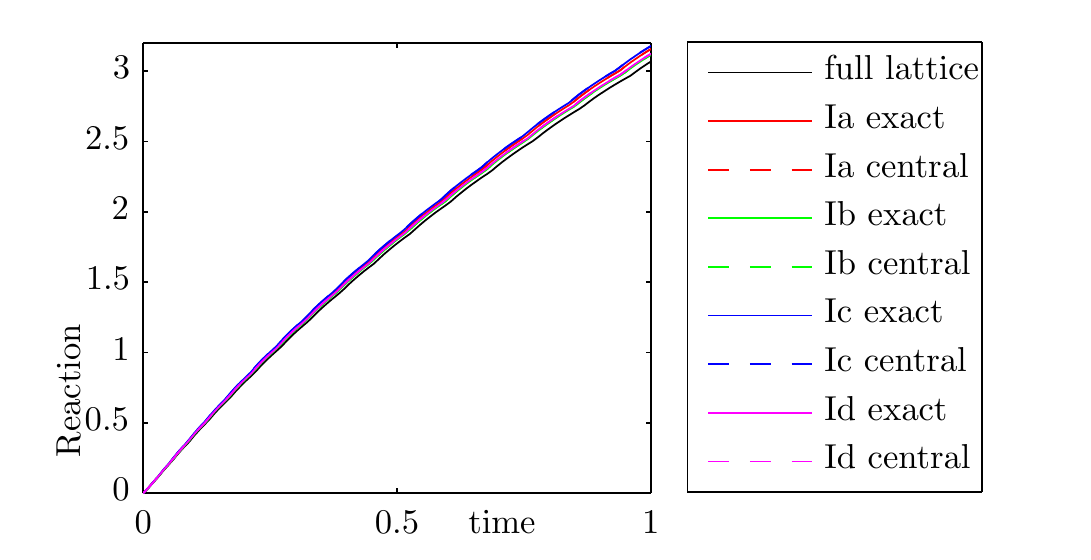}\label{SubSect:6.3:Fig:3b}}
		};
		\node[inner sep=0pt] (BreactZoom) at (5.4,1.55) {
			\includegraphics[scale=1]{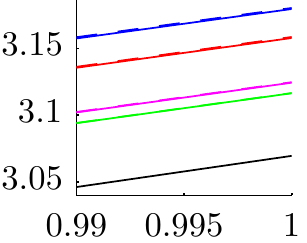}
		};

		\draw[black, thick, dashed, rounded corners] (9.75,2.8) rectangle (10.45,2);

		\draw[black] (3.8,0.3) rectangle (6.95,2.9);

		\draw[black, thick, dashed] (10.1,2.8) -- (6.95,2.9);
		\draw[black, thick, dashed] (10.1,2) -- (6.95,0.3);

	\end{tikzpicture}
	\caption{Results for the indentation test: (a)~close view of deformed configuration and the cumulative plastic slips in individual bonds for the full lattice solution, and~(b) the vertical reaction force of the indenter.}
	\label{SubSect:6.3:Fig:3}
\end{figure}

The energy evolution profiles are presented in Fig.~\ref{SubSect:6.3:Fig:4} for the exact, and in Fig.~\ref{SubSect:6.3:Fig:5} for the central summation rule. The results show that the energy balance~\eqref{E} holds in all cases, and that the differences between the two summation rules are practically negligible---as already suggested by Fig.~\ref{SubSect:6.3:Fig:3b}. Upon closer investigation (in Figs~\ref{SubSect:6.3:Fig:4b} and~\ref{SubSect:6.3:Fig:5b}), a better performance of the structured meshes compared to the unstructured ones can be observed again.
\begin{figure}
	\centering
	\begin{tikzpicture}
	\linespread{1}
	
	\node[inner sep=0pt] (energy) at (0,0) {
		\subfloat[exact summation rule]{\includegraphics[scale=1]{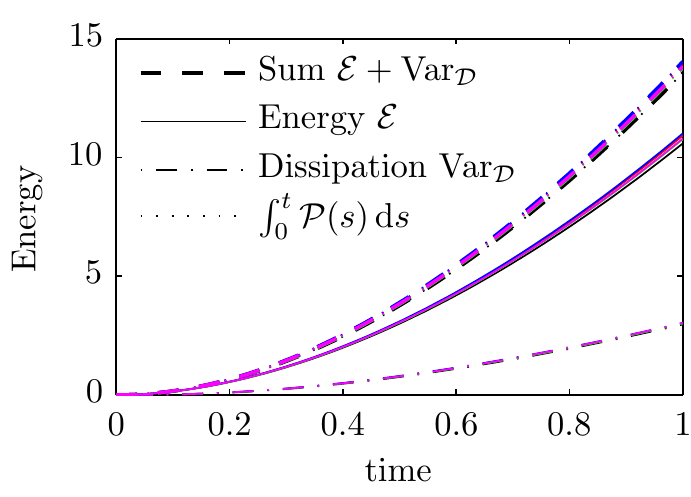}\label{SubSect:6.3:Fig:4a}}};
	\node[inner sep=0pt] (zoom) at (7.9,0.015) {
		\subfloat[zoom]{\includegraphics[scale=1]{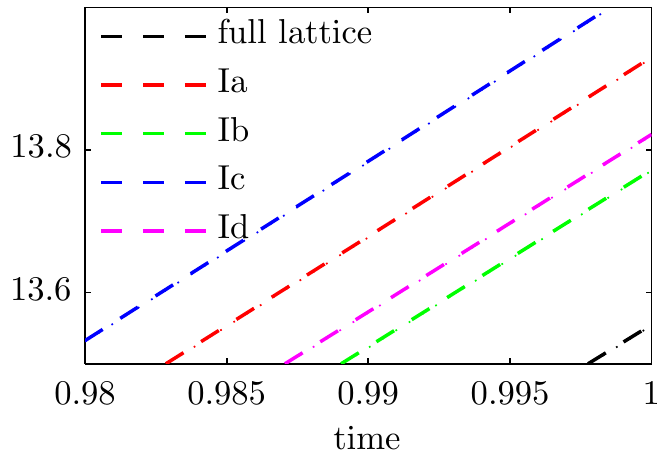}\label{SubSect:6.3:Fig:4b}}};
	
	\draw[black, thick, dashed, rounded corners] (2.2,1.5) rectangle (3.6,2.2);
	
	\draw[black, thick, dashed] (2.9,1.5) -- (5.4,-1.2);
	\draw[black, thick, dashed] (2.9,2.2) -- (5.4,2.4);
	
	\end{tikzpicture}
	\caption{Results for the indentation test: (a)~total energy evolution paths, (b)~zoom; the different meshes from Fig.~\ref{SubSect:6.3:Fig:2} using the exact summation rule are compared to the full-lattice solution.}
	\label{SubSect:6.3:Fig:4}
\end{figure}
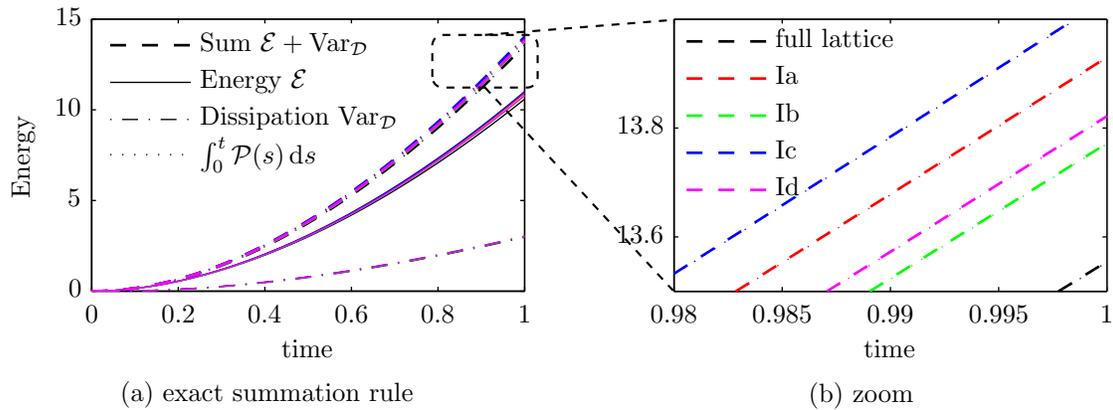
\begin{figure}
	\centering
	\begin{tikzpicture}
	\linespread{1}
	
	\node[inner sep=0pt] (energy) at (0,0) {
		\subfloat[central summation rule]{\includegraphics[scale=1]{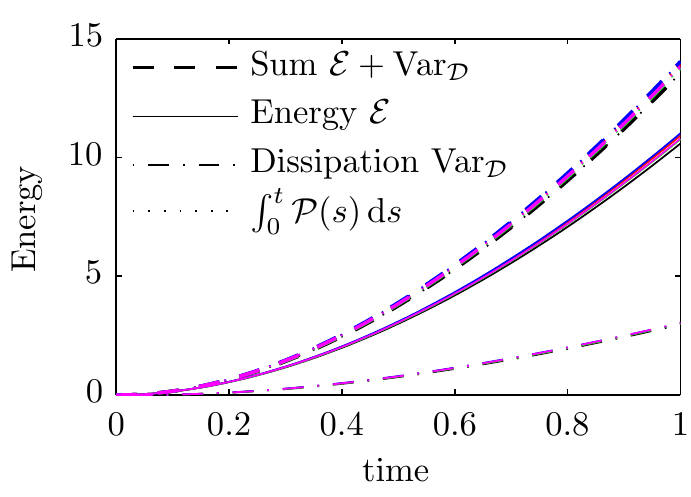}\label{SubSect:6.3:Fig:5a}}};
	\node[inner sep=0pt] (zoom) at (7.9,0.015) {
		\subfloat[zoom]{\includegraphics[scale=1]{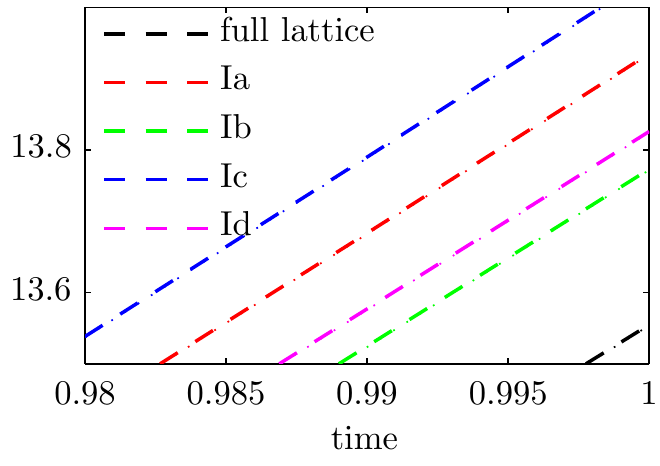}\label{SubSect:6.3:Fig:5b}}};
	
	\draw[black, thick, dashed, rounded corners] (2.2,1.5) rectangle (3.6,2.2);
	
	\draw[black, thick, dashed] (2.9,1.5) -- (5.4,-1.2);
	\draw[black, thick, dashed] (2.9,2.2) -- (5.4,2.4);

	\end{tikzpicture}
	\caption{Results for the indentation test: (a)~energy evolution paths, (b)~zoom; the different meshes from Fig.~\ref{SubSect:6.3:Fig:2} using the central summation rule are compared to the full-lattice solution.}
	\label{SubSect:6.3:Fig:5}
\end{figure}

Fig.~\ref{SubSect:6.3:Fig:6} shows the energy errors~$\varepsilon_{\widetilde{\Box}}$ of Eq.~\eqref{SubSect:6.1:Eq:1} for all systems. Here we see that the best agreement with the solution of the full lattice is obtained using regular fine mesh~$\mathrm{I}_\mathrm{b}$, whereas the largest error is obtained for unstructured coarse mesh~$\mathrm{I}_\mathrm{c}$. Because the total error is dominated by the interpolation, no significant correlation between the number of triangles with central sampling atoms that do not have all the neighbours in the same triangle (Fig.~\ref{SubSect:6.3:Fig:7b}) and the "Tot." error (Fig.~\ref{SubSect:6.3:Fig:6c}) is observed. Pearson's correlation coefficient amounts to~$-0.198$ in this case. For the summation error itself (i.e.~"Tot." minus~"Int." line in Fig.~\ref{SubSect:6.3:Fig:6c}), the correlation increases to~$0.266$, indicating only a mild dependence. By committing errors in energies of less than~$4\,\%$, we achieve a reduction up to a factor of~$50$ in terms of the number of degrees of freedom. The computational gain in terms of the number of sampling atoms is around a factor of~$40$.
\begin{figure}
	\centering
	\subfloat[$\varepsilon_\mathcal{E}$]{\includegraphics[scale=1]{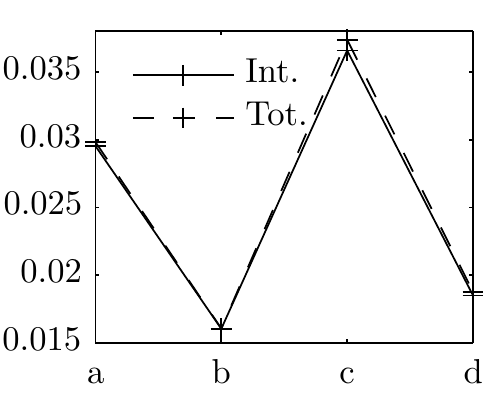}\label{SubSect:6.3:Fig:6a}}\hspace{0.5em}
	\subfloat[$\varepsilon_{\mathrm{Var}_\mathcal{D}}$]{\includegraphics[scale=1]{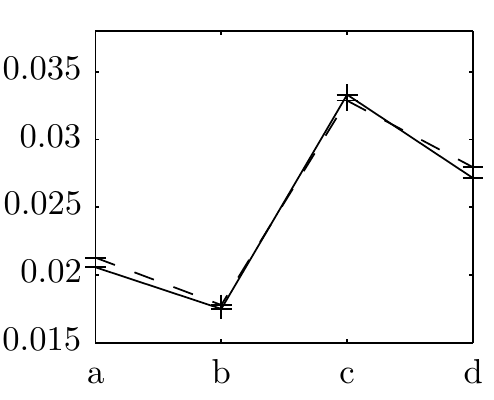}\label{SubSect:6.3:Fig:6b}}\hspace{0.5em}
	\subfloat[$\varepsilon_{\mathcal{E}+\mathrm{Var}_\mathcal{D}}$]{\includegraphics[scale=1]{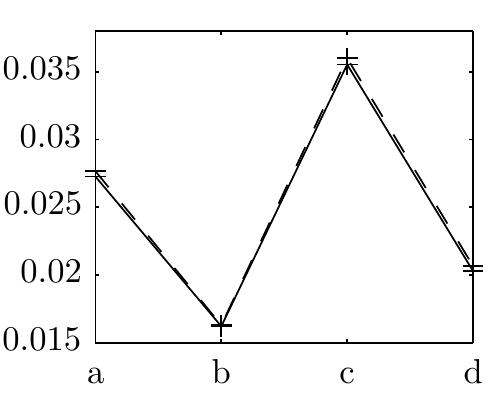}\label{SubSect:6.3:Fig:6c}}
	\caption{Results for the indentation test: relative errors~$\varepsilon_{\widetilde{\Box}}$ defined in Eq.~\eqref{SubSect:6.1:Eq:1} for the various meshes depicted in Fig.~\ref{SubSect:6.3:Fig:2}. "Int." relates to interpolation (exact summation rule) and "Tot." to interpolation plus summation (central summation rule).}
	\label{SubSect:6.3:Fig:6}
\end{figure}

Finally, Fig.~\ref{SubSect:6.3:Fig:7a} shows the local error~$\varepsilon_{z_\mathrm{p}}(t = 1)$ of Eq.~\eqref{SubSect:6.2:Eq8}. These results confirm that the size of the fully resolved region significantly influences the accuracy, whereas the choice of the summation rule has little effect on the local errors, except for mesh~$\mathrm{I}_\mathrm{c}$. Let us emphasize, however, that although the errors are acceptable (below~$14\,\%$), all QC approaches fail to identify in which bond the largest history variable occurs. This can be verified in Fig.~\ref{SubSect:6.3:Fig:3a}, where the bonds with the extreme plastic slips are presented by thick lines for the different systems. Note that the presented results correspond to the central summation rule only.
\begin{figure}
	\centering
	\subfloat[$\varepsilon_{z_\mathrm{p}}(t = 1)$]{\includegraphics[scale=1]{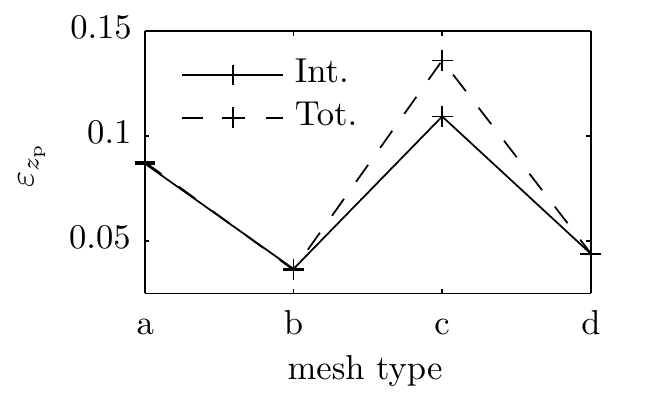}\label{SubSect:6.3:Fig:7a}}\hspace{0.5em}
	\subfloat[mesh characteristics]{\includegraphics[scale=1]{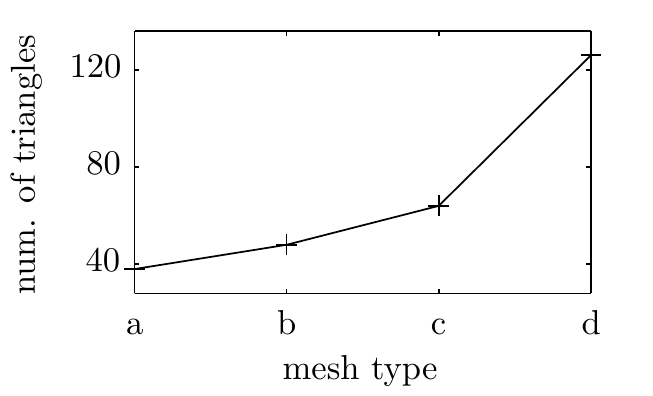}\label{SubSect:6.3:Fig:7b}}
	\caption{Results for the indentation test. (a)~Relative error~$\varepsilon_{z_\mathrm{p}}(t = 1)$ defined in Eq.~\eqref{SubSect:6.2:Eq8}. "Int." relates to interpolation (exact summation rule) and "Tot." to interpolation plus summation (central summation rule). (b)~The number of triangles for which the central sampling atoms do not have all their neighbours within the same triangle. The meshes from Fig.~\ref{SubSect:6.3:Fig:2} are used.}
	\label{SubSect:6.3:Fig:7}
\end{figure}
}
%
%
\section{Summary and Conclusions}
\label{Sect:6}
An analysis of the QuasiContinuum~(QC) approach for regular lattice systems with internal dissipative processes is presented based on the variational, energy-consistent formulation of~\cite{MieRou:2015}. The main results can be summarized as follows:
\begin{enumerate}
	\item The virtual-power-based QC method introduced by~\cite{BeexDisLatt} can be equivalently derived from an appropriate energy potential. As both QC schemes satisfy the energy balance, the variational structure of the virtual-power-based QC method is confirmed by this study.
	
	\item As a consequence of the central summation rule, the number of internal variables is highly reduced. An alternative approach may be to introduce geometrical constraints for the internal variables, possibly allowing a unique reconstruction of all unknowns (instead of only the kinematic variables). We have outlined briefly this perspective and emphasized the differences from the summation rule assumption.
	
	\item In the section describing the solution strategies, an alternating minimization method for nonsmooth potentials was specified for the QC framework.
	
	\item The example section has demonstrated the energy consistency of different QC schemes for \OR{three} benchmark examples involving uniform loading, pure bending, \OR{and indentation test}. In spite of large computational savings, all approximate solutions for \OR{all} examples proved to be in a good agreement with the full-lattice solution in terms of stored and dissipated energies (by accepting errors in energies only up to~$4\,\%$, the number of degrees of freedom may be reduced up to the factor of~$50$).
\end{enumerate}

The presented variational QC framework generalizes the original energy-based conservative quasicontinuum methodology for an entire class of rate-independent interactions and can easily be adjusted to incorporate e.g. damage phenomena merely by appropriate modifications of the energies. In addition, our results provide a convenient framework for adaptive variational QC methods in regular lattice networks with dissipative interactions, with potential applications in, e.g., fracture simulations of woven fabrics or polymers. All these topics are, nevertheless, beyond the scope of this contribution and will be reported separately.
%
\appendix
%
%
\section{Explicit Forms of Gradients and Hessians}
\label{Sect:A}
For the sake of completeness, we provide in this appendix the derivatives of~$\Pi_\mathrm{red}^k$, recall Eq.~\eqref{Sect:4:Eq:2}, with respect to~$\widehat{\bs{r}}$ explicitly. The internal force associated with atom~$\alpha$, $\bs{f}^\alpha_\mathrm{int}\in\mathbb{R}^{2\, n_\mathrm{ato}}$, reads as
\begin{equation}
\begin{aligned}
\bs{f}^\alpha_\mathrm{int}(\widehat{\bs{r}})=&\ 
\frac{\partial\widehat{\pi}^k_{\mathrm{red},\alpha}(\widehat{\bs{r}},\widehat{\bs{z}}_\mathrm{p};\bs{q}(t_{k-1}))}{\partial\widehat{\bs{r}}} = \frac{1}{2}\sum_{\beta\in B_\alpha}\frac{\partial\phi^{\alpha\beta}(\widehat{r}^{\alpha\beta},\widehat{z}_\mathrm{p}^{\alpha\beta})}{\partial\widehat{\bs{r}}^{\gamma}}=\\
=&\ \frac{1}{2}\sum_{\beta\in  B_{\alpha}}\phi'\frac{\widehat{\bs{r}}^{\alpha\beta}}{\widehat{r}^{\alpha\beta}}(\delta^{\beta\gamma}-\delta^{\alpha\gamma}),\ \gamma=1,\dots,n_\mathrm{ato},
\end{aligned}
\label{Sect:A:Eq:1}
\end{equation}
where~$\widehat{\pi}^k_{\mathrm{red},\alpha}$ denotes the reduced incremental site energy of atom~$\alpha$ with condensed variable~$\widehat{\bs{z}}_\mathrm{c}$, see also Eqs.~\eqref{SubSect:DissLatt:Eq:7b} and~\eqref{Sect:4:Eq:2}. The global force is then expressed as
\begin{equation}
\bs{f}(\widehat{\bs{r}})=\sum_{\alpha=1}^{n_\mathrm{ato}}\bs{f}^\alpha_\mathrm{int}(\widehat{\bs{r}}).
\label{Sect:A:Eq:Eq:2}
\end{equation}
Note that if external force vector~$\bs{f}_\mathrm{ext}$ is present, it is simply subtracted from the right-hand-side of Eq.~\eqref{Sect:A:Eq:Eq:2}. The stiffness matrix associated with an atom site~$\alpha$, $\bs{K}^\alpha\in\mathbb{R}^{2\, n_\mathrm{ato}}\times\mathbb{R}^{2\, n_\mathrm{ato}}$, reads as
\begin{equation}
\begin{aligned}
\bs{K}^{\alpha}(\widehat{\bs{r}}) =&\ \frac{\partial^2\widehat{\pi}_\mathrm{red,\alpha}^k(\widehat{\bs{r}},\widehat{\bs{z}}_\mathrm{p};\bs{q}(t_{k-1}))}{\partial\widehat{\bs{r}}\partial\widehat{\bs{r}}}=
\frac{1}{2}\sum_{\beta\in B_\alpha}\frac{\partial^2\phi^{\alpha\beta}(\widehat{r}^{\alpha\beta},\widehat{z}_\mathrm{p}^{\alpha\beta})}{\partial\widehat{\bs{r}}^{\gamma}\partial\widehat{\bs{r}}^{\delta}}=\\
 =&\ \frac{1}{2}\sum_{\beta\in B_\alpha}\left[\frac{\phi'}{\widehat{r}^{\alpha\beta}}\delta_{mn}+\left(\frac{\phi''}{(\widehat{r}^{\alpha\beta})^2}-\frac{\phi'}{(\widehat{r}^{\alpha\beta})^3}\right)\widehat{\bs{r}}^{\alpha\beta}\otimes\widehat{\bs{r}}^{\alpha\beta}\right](\delta^{\beta\gamma}-\delta^{\alpha\gamma})(\delta^{\beta\delta}-\delta^{\alpha\delta}),\\
&\ m,n=1,2,\ \gamma,\delta=1,\dots,n_\mathrm{ato},
\end{aligned}
\label{Sect:A:Eq:3}
\end{equation}
whereas global stiffness is expressed as
\begin{equation}
\bs{K}(\widehat{\bs{r}})=\sum_{\alpha=1}^{n_\mathrm{ato}}\bs{K}^\alpha(\widehat{\bs{r}}).
\label{Sect:A:Eq:4}
\end{equation}
Here, we have used the relation
\begin{equation}
\frac{\partial\widehat{r}^{\alpha\beta}}{\partial\widehat{r}_m^\gamma}=\frac{\widehat{r}_m^{\alpha\beta}}{\widehat{r}^{\alpha\beta}}(\delta^{\beta\gamma}-\delta^{\alpha\gamma}),\ m=1,2,
\label{Sect:A:Eq:5}
\end{equation}
and for brevity we have denoted
\begin{equation}
\phi'=\frac{\partial\phi^{\alpha\beta}(\widehat{r}^{\alpha\beta},\widehat{z}_\mathrm{p}^{\alpha\beta})}{\partial \widehat{r}^{\alpha\beta}},\ \phi''=\frac{\partial^2\phi^{\alpha\beta}(\widehat{r}^{\alpha\beta},\widehat{z}_\mathrm{p}^{\alpha\beta})}{\partial(\widehat{r}^{\alpha\beta})^2}.
\label{Sect:A:Eq:6}
\end{equation}
Above, $m,n$ relate to spatial dimensions, $\alpha,\beta$ relate to atoms, $\delta_{mn}$ denotes the Kronecker-delta product with respect to spatial coordinates, $\delta^{\alpha\beta}$ denotes the Kronecker-delta product with respect to atoms, and~$\bs{a}\otimes\bs{b}=a_mb_n$ denotes the tensor product of vectors~$\bs{a}$ and~$\bs{b}$.
%
%
%
%
\section*{Acknowledgements}
Financial support for this work from the Czech Science Foundation (GA\v{C}R) under project No.~14-00420S is gratefully acknowledged.

%
%


\end{document}